\documentclass[letterpaper, 10 pt, conference]{ieeeconf}  %

\IEEEoverridecommandlockouts                              %

\usepackage{amsthm}
\usepackage{hyperref}

\title{\LARGE \bf
 Designing Inferable Signaling Schemes for Bayesian Persuasion
}

\author{Caleb Probine$^{1}$, Mustafa O. Karabag$^{1}$, Ufuk Topcu$^{1}$%
\thanks{$^{1}$All authors are with the University of Texas at Austin.}%
}

\usepackage{xcolor}

\usepackage{amssymb}
\usepackage{amsmath}
\usepackage{xcolor}
\usepackage{xargs}
\usepackage{tikz}
\usepackage{cancel}
\usepackage{comment}
\usepackage{nicefrac}
\usetikzlibrary{patterns,arrows,arrows.meta,calc,shapes,shadows,decorations.pathmorphing,decorations.pathreplacing,automata,shapes.multipart,positioning,shapes.geometric,fit,circuits,trees,shapes.gates.logic.US,fit,backgrounds,decorations.text,topaths,shadings,external,calc,arrows.meta}

\newcommandx{\mk}[2][1=inline]{\todo[linecolor=blue,backgroundcolor=blue!25,bordercolor=blue,#1]{\scriptsize{[MK]~ #2}}}
\newcommandx{\cp}[2][1=inline]{\todo[linecolor=green,backgroundcolor=green!25,bordercolor=green,#1]{\scriptsize{[CP]~ #2}}}

\newcommand{\post}[1]{y_{#1}^{\pi}}
\newcommand{\postEst}[1]{\hat{y}_{#1}^{\pi}}
\newcommand{\genPost}[2]{{#1}_{#2}}
\newcommand{\genPostEst}[3]{\hat{#1}_{#2}^{#3}}

\newcommand{\actDist}[1]{\calD_{\hat{a}_k(#1)}}
\newcommand{\sigProb}[1]{p^\pi(#1)}

\newcommand{\calD}{\mathcal{D}}
\newcommand{\calA}{\mathcal{A}}
\newcommand{\calO}{\mathcal{O}}
\newcommand{\calX}{\mathcal{X}}

\newcommand{\ie}{i.e.}

\newcommand{\sg}{s}
\newcommand{\SG}{S}

\newcommand{\uvec}[2]{u_{#1}(:,#2)}
\newcommand{\uvecsup}[3]{u_{#1}^{#3}(:,#2)}

\newcommand{\stLeadSet}{A^L}
\newcommand{\stFolSet}{A^F}

\newcommand{\psgdvar}{X}

\newtheorem{proposition}{Proposition}
\newtheorem{corollary}{Corollary}
\newtheorem{definition}{Definition}
\newtheorem{lemma}{Lemma}
\newtheorem{theorem}{Theorem}
\newtheorem{problem}{Problem}
\newtheorem{fact}{Fact}

\newenvironment{proof_alt}[1][Proof]{%
  \par\noindent\textit{#1:} }{\hfill$\blacksquare$\par}

\newif\iflong
 \longtrue   %

\newif\ifappfn
\appfntrue  %

\begin{document}

\maketitle
\thispagestyle{empty}
\pagestyle{empty}

\setcounter{footnote}{1}

\begin{abstract}
In Bayesian persuasion, an informed sender, who observes a state, commits to a randomized signaling scheme that guides a self-interested receiver's actions.
Classical models assume the receiver knows the commitment.
We, instead, study the setting where the receiver infers the scheme from repeated interactions.
We bound the sender’s performance loss relative to the known-commitment case by a term that grows with the signal space size and shrinks as the receiver’s optimal actions become more distinct.
We then lower bound the samples required for the sender to approximately achieve their known-commitment performance in the inference setting. 
We show that the sender requires more samples in persuasion compared to the leader in a Stackelberg game, which includes commitment but lacks signaling. 
Motivated by these bounds, we propose two methods for designing inferable signaling schemes, one being stochastic gradient descent (SGD) on the sender’s inference-setting utility, and the other being optimization with a boundedly-rational receiver model.
SGD performs best in low-interaction regimes, but modeling the receiver as boundedly-rational and tuning the rationality constant still provides a flexible method for designing inferable schemes.
Finally, we apply SGD to a safety alert example and show it to find schemes that have fewer signals and make citizens’ optimal actions more distinct compared to the known-commitment case.
\footnote{\label{fn:extended_version}We give extended proofs of all results in the appendix.}
\end{abstract}

\section{Introduction}

Bayesian persuasion models interactions between a sender and a receiver with potentially misaligned utilities \cite{kamenica2011bayesian}.
The sender observes the state of an environment, from which they generate signals, which the receiver, in turn, uses to choose actions.
Bayesian persuasion has applications ranging from managing congestion \cite{das2017reducing} to autonomous driving \cite{peng2019bayesian}.

The sender seeks the optimal signaling scheme, i.e., a mapping from states to signal distributions, to commit to.
On seeing a signal, a receiver who knows the scheme constructs a posterior on the state and takes a utility-maximizing action.
The sender maximizes their expected utility under the state-action distribution the receiver's behavior induces.

We study persuasion in the \emph{inference} setting, where the sender and receiver interact in a sequence of rounds, over which the receiver learns the scheme.
At each round, the receiver reacts to a signal using an estimate of the scheme, and then observes the true state, updating the estimate accordingly.
This model contrasts \emph{known-commitment} settings, where the receiver knows the scheme the sender commits to.

Bayesian persuasion in the inference setting is natural in many applications.
For example, consider a planner designing safety alerts.
The planner aims to ensure people avoid unsafe areas while taking into account their goals.
The planner must design schemes that simultaneously incentivize safe behavior while being easy for people to learn.

Commitment in Bayesian persuasion mirrors %
Stackelberg games \cite{conitzer2006computing}, where a leader commits to a strategy while predicting a follower's reaction.
However, the sender in Bayesian persuasion additionally has privileged information.

While prior work studies Stackelberg games where followers learn leaders' commitments from samples \cite{an2012security,blum2014lazy,muthukumar2020learning,karabag2023should}, such approaches do not extend to persuasion.
In the Stackelberg case, the follower must only learn a single distribution over actions, whereas in persuasion, the receiver must learn a joint distribution over states and signals.

In this work, we first study how schemes designed in the known-commitment setting degrade under inference.
We bound the gap in the sender's utility loss between the two settings by a term that decreases with the conditional entropy of the state given the signal, increases with the number of signals, and increases as the receiver's posterior approaches decision boundaries.
Thus, the sender must balance conflicting objectives of 
hiding information according to their utility and providing information to promote inferability.

We then 
lower bound the samples needed in the inference setting so that the sender can approximately achieve their known-commitment utility.
We give a sequence of persuasion problems that vary in the number of states and a sample lower bound as a function of this number of states.
We compare these lower bounds to upper bounds for an analogous series of Stackelberg games that vary in the number of leader actions and conclude that, from an inferability perspective, persuasion problems have higher sample complexity.

We finally propose two approximate methods for designing inferable signaling schemes.
The naive solution is to optimize over the distribution of scheme estimates for the receiver, but this approach leads to a non-convex program with size exponential in the number of states  
due to the size of the set of estimates.
We propose a stochastic approximation method that overcomes the need to sample from the large number of distribution estimates by sampling directly from the distribution on the receiver's actions.
We also propose a regularization method that solves the persuasion problem under the assumption of bounded rationality receivers.

While both approaches still require solving non-convex optimization problems, we show that the computed schemes provide significant improvements on the naive baseline of using known-commitment-optimal signaling schemes in the inference setting. We then apply the stochastic approximation method to the aforementioned safety-alert example.

\subsection{Related work}

\subsubsection{Signaling for information-limited receivers}

Various work studies signaling with learning receivers.
For example, assuming the receiver is a no-regret learner, prior work characterizes the reward the sender can attain when the receiver does not know the state distribution \cite{camara2020mechanisms, collina2024efficient} or does not know the signaling scheme \cite{lin2024generalized,lin2024information}.
Another model involves the sender providing state distributions to the receiver, who trusts these distributions if they pass consistency tests against historical data \cite{jain2024calibrated}.
In contrast to these works, we
study a specific model with a receiver that maximizes utility against an empirical estimate of the scheme, as in fictitious play \cite{fudenberg1998theory}.

Robust signaling schemes also address the issue of limited-information receivers.
For example, one can design schemes for a receiver lacking information about the state distribution
by ensuring schemes satisfy incentive compatibility constraints for sets of distributions \cite{zu2021learning} or allowing the receiver to take sub-optimal actions \cite{camara2022mechanism}.
However, these methods still assume the receiver knows the scheme.
More generally, various work explores signaling for different receiver models \cite{yang2024computational,tang2021bayesian} including bounded rationality \cite{feng2024rationality}.
We study a particular model where the receiver takes optimal actions with respect to an empirical estimate of the signaling strategy.

Existing work also studies signaling when the sender lacks perfect commitment \cite{min2021bayesian}, and, for example, must implement commitment via reputation \cite{best2024persuasion}.
However, the studied settings explore equilibrium behavior in dynamic games and thus assume the receiver is rational with respect to the true scheme.
We allow sender commitment, and model the receiver as reacting rationally to empirical estimates of this commitment.

Finally, information limitations for the receiver arise via communication constraints, e.g., when the signaling channel is bit-limited \cite{dughmi2016persuasion,le2019persuasion}.
However, such settings still allow the receiver to observe the committed strategy.
While we do not impose channel constraints, we note that the results we give imply that schemes with few signals lead to high sender reward under inference.
While recent work on constrained signaling motivates the problem via explainability \cite{chen2025explainable}, the underlying models still assume known-commitment.

\subsubsection{Empirical inference in repeated games}

In the inference setting, the receiver maximizes utility using an empirical estimate of the scheme, as in fictitious play \cite{fudenberg1998theory}, and previous work characterizes repeated play with agents using fictitious play in perfect information games \cite{dong2022optimal,assos2024maximizing,vundurthy2023intelligent}.
However, these works allow the ego player to change their strategy, while we require the sender to commit.

When the leader commits in a Stackelberg game, and the follower infers this commitment from samples, various work characterizes the effect of this learning on the sender's utility \cite{an2012security,blum2014lazy,muthukumar2020learning,karabag2023should}.
However, such characterizations do not extend to persuasion.
We note that while \cite{muthukumar2020learning} discusses a persuasion example, the methods do not provide techniques for general persuasion problems.
The framework we study is closest to \cite{karabag2023should} as we consider the same interaction structure, with the addition of signals and hidden information.

\section{Preliminaries}
For $m$ in $\mathbb{N}$, $[m]$ is the set $\{1,\ldots,m\}$.
The set of distributions on set $\calA$ is $\Delta^\calA$. 
The binomial distribution is $\mathsf{Bin}(k,p)$ and the multinomial distribution is $\mathsf{Multi}(k,q)$.
For $x,y \in \mathbb{R}^n$, $\langle x,y \rangle = \sum_{i \in [n]} x_i y_i$ is the inner-product, and $||x||_2 = \smash{\langle x,x \rangle^{\nicefrac{1}{2}}}$ is the $l^2$ norm.
For a distribution $x\in \Delta^{[n]}$,  $\nu(x) = \smash{(\sum_{i \in [n]} x_i(1-x_i) )^{\nicefrac{1}{2}}}$ is the stochasticity of $x$.
The all ones vector is $\mathbf{1}$ and the all zeros vector is $\mathbf{0}$.
For $p,q \in \Delta^{[n]}$, the Kullback-Leibler (KL) divergence is $D_{KL}(p||q) = \sum_{i\in[n]} p_i \log (p_i/q_i)$.
The entropy of a variable with distribution $p \in \Omega^\calX$ is
$
    H(p) = -\sum_{x\in \calX} p(x)\log(p(x)).
$
For two random variables with joint distribution $p(x,y)$,  $H(x|y) = \sum_{y} p(y) \sum_{x} p(x|y) \log(p(x|y))$ is their conditional entropy. %

\textbf{Bayesian Persuasion. }
Bayesian-persuasion models interactions between a sender, who observes a state $\omega$ from a state space $\Omega$ drawn according to $\mu \in \Delta^\Omega$, and a receiver who takes an action $a$ from set $A$.
The sender and receiver have respective utilities 
$u_S: \Omega \times A \rightarrow \mathbb{R}$
and
$u_R: \Omega \times A \rightarrow \mathbb{R}$.
The vector in $\mathbb{R}^\Omega$ that defines $u_{S}$ for action $a$ is $\uvec{S}{a}$.
We define $\uvec{R}{a}$ analogously.
A signaling scheme is a mapping $\pi : \Omega \rightarrow \Delta^\SG$, for a signal set $\SG$.
We use $\pi(\omega,\sg)$ to denote $p^\pi(\sg|\omega)$
The marginal probability of signal $\sg$ is $\sigProb{\sg}$.
In the known-commitment setting, the game evolves as follows.
\begin{enumerate}
    \item The sender commits to a strategy
    $\pi : \Omega \rightarrow \Delta^\SG$.
    \item The sender observes the state $\omega \sim \mu$.
    \item The sender draws a signal $\sg \sim \pi(\omega)$.
    \item The receiver computes the posterior $\post{\sg} \in \Delta^\Omega$ with
    \begin{equation}
        \post{\sg} := p^\pi(\omega|\sg) = \frac{\pi(\omega,\sg) \mu(\omega)}{\sum_{\omega' \in \Omega} \pi(\omega',\sg) \mu(\omega')},
    \end{equation}
    and chooses an action $a$ in  
    \begin{equation}
        \arg \max_{a \in A} \sum\nolimits_{\omega \in \Omega} u_R(\omega,a) p^\pi(\omega|\sg).
    \end{equation}
\end{enumerate}
When $\arg \max_a \langle \uvec{R}{a}, \post{\sg} \rangle$ is not a singleton, the receiver takes an action maximizing $\langle \uvec{S}{a}, \post{\sg} \rangle$, as is common in persuasion \cite{dughmi2016algorithmic}.
When utilities are clear from context, $a(x)$ is the action the receiver takes for state distribution $x$, and $E_a = \{y \in \Delta^\Omega| a(y) = a\}$ is the set of distributions where the receiver takes action $a$.
On observing $\sg$, the receiver's posterior is $y_{\sg}^\pi$, and they take action $a(\post{\sg})$. 
We use $a^\pi(\sg)$ to denote this action.
The sender maximizes expected utility
\begin{equation}
    BPR(\pi) := \mathbb{E}_{\omega \sim \mu, \sg \sim \pi(\omega)}[u_S(\omega,a^\pi(\sg)].
\end{equation}

\textbf{Stackelberg games. }
Stackelberg games model interactions between a leader who commits to a strategy, \ie,  a distribution over actions $\stLeadSet$, and a follower who knows this commitment and takes actions in a set $\stFolSet$.
The leader and follower have utility functions $u_L:\stLeadSet \times \stFolSet \rightarrow \mathbb{R}$ and $u_F : \stLeadSet \times \stFolSet \rightarrow \mathbb{R}$ respectively.
On observing the leader's commitment, the follower takes a utility maximizing action, where we again break ties in favor of the leader \cite{conitzer2006computing}.

\section{Setting and Problem Statement}

The known-commitment model relies on the receiver knowing the signaling scheme. 
We study the \emph{inference} setting where the receiver learns the scheme from past interactions.

\begin{figure}
\centering

\includegraphics[width=1.0\linewidth]{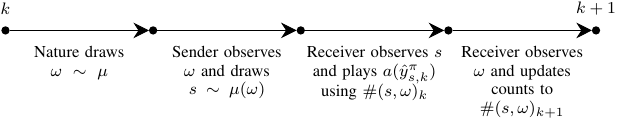}
\caption{\textbf{Bayesian persuasion in the inference setting.}
At each round, the receiver takes action using their empirical estimate of the scheme, and then updates the estimate using the observed state.}
\label{fig:timeline}
\end{figure}

In the inference setting, the sender and receiver interact over a series of rounds, and the receiver incrementally estimates the scheme, as shown in Figure~\ref{fig:timeline}.
Fix a scheme $\pi$.
At round $k$, for a signal $\sg \in \SG$, the receiver has an estimate $\hat{y}^\pi_{\sg,k}$ of the posterior.
On seeing $\sg$, the receiver takes the action $a(\hat{y}^\pi_{\sg,k})$ their posterior estimate induces.
The receiver then observes the true state $\omega$ and updates their estimates.

The sender's reward $IR_k$ at round $k$ is the expected reward when accounting for the receiver's estimation. 
That is,
\begin{equation}
    IR_k(\pi) = \mathbb{E}_{\omega \sim \mu, \sg \sim \pi(\omega)}\mathbb{E}_{a \sim \actDist{\sg}} [u_S(\omega, a)]
\end{equation}
where $\actDist{\sg}$ is the distribution on the action that the receiver takes at round $k$ on seeing signal $\sg$ using their estimate. %

We first investigate how schemes designed for the known-commitment setting perform under inference.
\begin{problem}
    \label{prob:boundProblem}
    Let a scheme $\pi$ and a Bayesian persuasion problem $(\Omega, A, u_S, u_R, \mu)$ be given. 
    For a fixed $k$, what is the value of $IR_k(\pi) - BPR(\pi)$? Conversely, given some $\epsilon$, what is the minimum $k$ such that $IR_k(\pi) \geq BPR(\pi) - \epsilon$?
\end{problem}

We then study the design of inferable schemes.
\begin{problem}
    \label{prob:optProblem}
    Given a Bayesian persuasion problem, how can we solve the optimization problem $\max_{\pi} IR_k(\pi)$?
\end{problem}
\noindent We remark that our methods extend to the problem of maximizing cumulative reward, \ie, $\sum_{j=1}^k IR_j(\pi)$.

These problems are ill-defined for arbitrary estimation mechanisms, and we consider receivers who maintain counts of each $(\omega,\sg)$ pair.
The receiver computes their posterior as
\begin{equation}
    \label{eq:postEstPS}
    \postEst{\sg,k}(\omega) = \frac{\# (\omega,\sg)_k }{\sum_{\omega \in \Omega} \# (\omega,\sg)_k},
\end{equation}
where $\#(\omega,\sg)_k$ is the number of occurrences of $(\omega,\sg)$ before round $k$.
We consider a setting where, before the first round, the receiver observes a sample from each posterior $\post{\sg}$, so that \eqref{eq:postEstPS} is well-defined.
Letting $p^\pi \in \Delta^{\Omega \times \SG}$ be the joint state-signal distribution, the counts have the following distribution.
\begin{multline}
    \#(\omega,\sg)_k = Z(\omega,\sg) + Y_{\sg}(\omega) : \\
     Z \sim \mathsf{Multi}(k-1,p^\pi) \quad Y_{\sg} \sim \mathsf{Multi}(1,\post{\sg}).
\end{multline}
We encode the initial sample from $\post{\sg}$ using $Y_{\sg}$ while $Z$ encodes that,
at each round, the receiver sees one signal-state pair, distributed according to $p^\pi$.
For each signal, the distribution on $\#(\omega,\sg)_k$ induces a distribution $\calD_{\postEst{\sg,k}}$ on posterior estimates, which induces the distribution $\actDist{\sg}$.

\section{Bounding value-loss under inference}
\label{sec:GapUpperBound}

We upper bound the gap between the sender's value in the known-commitment and inference settings in terms of the number $k$ of samples and the properties of the scheme $\pi$.
We demonstrate that schemes with posteriors that are far from decision boundaries and less stochastic have smaller gaps, which is consistent with the gap bound for the Stackelberg case \cite{karabag2023should}.
Contrasting the Stackelberg case, the bound we give additionally depends on the scheme's properties via the marginal probability $\sigProb{\sg}$ with which each signal $\sg$ is sent.

The upper bound depends on 
the distance between the signaling scheme and the receiver's decision boundary, and so we define an appropriate distance.
Let 
$$C_{a,a'} = \{y\in \Delta^\Omega| \langle y, \uvec{R}{a} - \uvec{R}{a'} \rangle = 0\},$$
\ie \ those distributions for which actions $a$ and $a'$ have equal value to the receiver.
For a distribution $x$ in $E_a$, 
$$
    d(x) = \min_{a' \in A \setminus\{a\}} \min_{y \in C_{a,a'}} ||y - x||_2
$$
is the distance of $x$ to the nearest boundary that defines $E_a$.

\begin{proposition}
    \label{prop:upperBoundMain}
    If $\mathsf{Range}(u_S) \subset [0,1]$, then
    \begin{equation}
        \mathsf{GAP}_\pi := BPR(\pi) - IR_k(\pi) \leq  \sum_{\sg \in \SG} \sqrt{\frac{\sigProb{\sg}}{k}} \frac{\nu(\post{\sg})}{d(\post{\sg})}.
    \end{equation}
\end{proposition}
\textit{Proof sketch:}
\iflong
\footref{fn:extended_version}
\fi
We bound the probability that, for signal $\sg$, the receiver chooses an action different from $a^\pi(\sg)$, and bounding this probability via Markov's inequality, we have
\begin{equation}
    \mathsf{GAP}_\pi \leq \sum\nolimits_{\sg\in\SG} \sigProb{\sg} \mathbb{E}[||\postEst{\sg,k} - \post{\sg}||_2]({1}/{d(\post{\sg})}).
\end{equation}
We bound the expected distance by conditioning on the number $\# (\sg)_k$ of times  the receiver sees signal $s$, \ie
\begin{multline}
    \sum_{h = 1}^k p(\# (\sg)_k = h) \mathbb{E}[||\postEst{\sg,k} - \post{\sg}||_2| \# (\sg)_k = h]
    \\ \leq \nu(\post{\sg}) \sum_{h = 1}^k p(\# (\sg)_k = h)  \sqrt{\frac{1}{h}}
    \leq \nu(\post{\sg}) \sqrt{\frac{1}{\sigProb{\sg}}}.
\end{multline}
The first inequality uses Lemma 1 in \cite{karabag2023should}, and the second inequality uses the binomial distribution of $\# s$ and the properties of binomial moments \cite{chao1972negative}.
$\square$

Compared to the gap bound in \cite{karabag2023should}, Proposition~\ref{prop:upperBoundMain} differs in the $\sqrt{\sigProb{\sg}}$ term, which encodes that signaling schemes with many equiprobable signals may be hard to infer.
The bound may initially suggest that increasing $\sigProb{\sg}$ increases the gap. 
However, we note that signal probabilities $\sigProb{\sg}$ are constrained, as $(\sigProb{\sg})_{\sg \in \SG}$ lies on the simplex. Hence, $\sum_{\sg \in \SG} \sqrt{\sigProb{\sg}}$ is largest when the marginal distribution on signals is uniform.
If $\nicefrac{\nu(\post{\sg})}{d(\post{\sg})}$ is constant in $\sg$, and $(\sigProb{\sg})_{\sg \in \SG}$ is uniform, then $BPR(\pi) - IR_k(\pi) \leq \sqrt{\nicefrac{|\SG|}{k}}$. 
That is, $\mathsf{GAP}_\pi$ will decrease as the signal space gets smaller.

Proposition~\ref{prop:upperBoundMain} implies bounds on the samples required for the sender to approximately achieve their optimal known-commitment value in the inference setting.
Given $\pi$ such that $BPR(\pi) \geq \max_{\pi'} BPR(\pi') - \epsilon/2$,
\begin{equation}
    k \geq \calO\left( \nicefrac{1}{ \left(\sum_{\sg \in \SG} \sqrt{\sigProb{\sg}} \frac{\nu(\post{\sg})}{d(\post{\sg})}\right)^2} \cdot \nicefrac{1}{\epsilon^2}\right)
\end{equation}
implies that $IR_k(\pi) \geq \max_{\pi'} BPR(\pi') - \epsilon$.

We additionally have the following information-theoretic bound on the gap in value between these two settings.
\begin{corollary}
\label{cor:MIBound}
Let  $d(\pi) = \min_{\sg} d(\post{\sg})$. If $\mathsf{Range}(u_S) \subset [0,1]$,
\begin{equation}
    \label{eq:MISigBound}
    \mathsf{GAP}_\pi \leq \frac{1}{ d(\pi)} \sqrt{|\SG|} \sqrt{H(\omega|\sg)}.
\end{equation}
\end{corollary}
This result is looser than Proposition~\ref{prop:upperBoundMain} and follows by first bounding stochasticity by entropy for each posterior, and then applying H\"older's inequality to $\sum_{\sg\in \SG} \sqrt{\sigProb{\sg} H(\post{\sg})}$.

We can cast signaling as a rate-distortion problem \cite{cover1999elements} via Corollary~\ref{cor:MIBound}. 
Indeed, known-commitment-optimal schemes may limit the receiver's information, but to decrease the bound in \eqref{eq:MISigBound}, the sender can decrease the conditional entropy of the state given the signal and reveal information.

\section{Sample complexity lower bounds}

We now lower bound the number of samples so that the sender's reward can approach $\max_{\pi'} BPR(\pi')$ in the inference setting.
We construct a series of games, defined by the number $n$ of states, and for each game, we lower bound the number $k$ of samples so that $IR_k(\pi) \geq \max_{\pi'} BPR(\pi') - \epsilon$ for any scheme $\pi$ such that $BPR(\pi) \geq \max_{\pi'} BPR(\pi') - \epsilon$.

Previous work provides lower bounds for a single fixed-size Stackelberg game \cite{karabag2023should}.
In contrast, we provide examples for general game dimensions and account for the signaling component present in Bayesian persuasion.

We compare Bayesian persuasion lower bounds to upper bounds for a related series of Stackelberg games and show that Bayesian persuasion problems have a higher sample complexity than Stackelberg games in the inference setting.

\subsection{A game that requires many equiprobable signals}

\begin{figure}
    \centering
    \includegraphics[width=.65\linewidth]{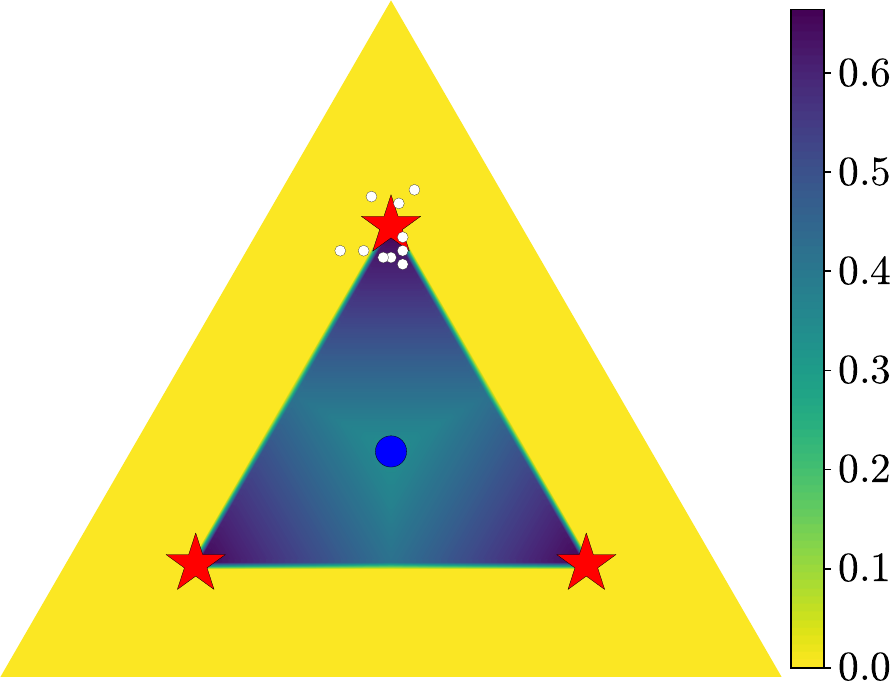}
    \caption{
    \textbf{The optimal scheme for a flower game $G^{\nicefrac{1}{6},3}$ induces posteriors near to decision boundaries.}
    Red stars are posteriors $\post{\sg}$, the blue dot is the prior, and 
    white dots are empirical distributions sampled from a posterior.
    We color the simplex by $f(y) = \sum_{\omega \in \Omega} u_S(\omega,a(y)) y(\omega)$. 
    For a signal $\sg$, the leader's  known-commitment expected reward, conditional on the signal, is $f(\post{\sg})$. 
    The known-commitment-optimal posteriors lie in regions where $f(y)$ is large.
    For posteriors at the inner triangle's edge, the receiver is likely to take actions in the inference setting that give the sender zero reward.
   }
    \label{fig:rewardViz}
\end{figure}

We design a sequence of Bayesian persuasion problems defined by the number $n$ of states, and a parameter $\tau$ that controls the geometry of the problem.
\begin{definition}[Flower game]
    For $n \in \mathbb{N}$ and $ \tau \in [0,\nicefrac{1}{n}]$,
    the flower game $G^{\tau,n}$ is the persuasion 
    problem that the tuple $(\Omega_n,A_n,u_S^{\tau,n}, u_R^{\tau,n}, \mu_n)$ defines.
    The set $\Omega_n$ of states is $[n]$, where $\mu_n(i) = \nicefrac{1}{n}$ for all $i \in [n]$.
    The set of actions is 
    $$A_n = \{a_j\}_{j=1}^n \cup \bigcup\nolimits_{j \in [n]} \{a_{(jk)}\}_{k \in [n]\setminus j} .$$
    The utility $u_R^{\tau,n}$ satisfies
    $
        \uvecsup{R}{a_j}{\tau,n}= e_j %
    $
    for $j$ in $[n]$ and
    \begin{equation}
         \uvecsup{R}{a_{(jk)}}{\tau,n} = e_j + \tau \mathbf{1} - e_k \text{ for } j \in [n], k \in [n] \setminus \{j\}.
    \end{equation}
    The utility $u_S^{\tau,n}$ satisfies $\uvecsup{S}{a_j}{\tau,n} = e_j$ for $j$ in $[n]$ and 
    \begin{equation}
         \uvecsup{S}{a_{(jk)}}{\tau,n} = \mathbf{0}, \text{ for } j \in [n], k \in [n] \setminus \{j\}.
    \end{equation}
\end{definition}

Intuitively, the receiver takes action $a_i$ when state $i$ is most likely, but all states have probability at least $\tau$. Lemma~\ref{lem:actOptCharac} characterizes these cases where the receiver takes action $a_i$.\iflong\footref{fn:extended_version}\fi
\begin{lemma}
    \label{lem:actOptCharac}
    For $y$ a distribution on states, $a_i \in \arg\max_{a\in A} \langle \uvec{R}{a}, y\rangle$
    if and only if 1) $y_i \geq y_j $ for all $j \neq i$, and 2) $y_j \geq \tau$ for all $j \neq i$.
\end{lemma}

In the optimal scheme $\pi^*_n$, there are $n$ signals that are sent with equal marginal probability, and the posterior for signal $i$ is the distribution that maximizes the probability of state $i$, for all distributions that induce action $a_i$.
\begin{fact}
    \label{fac:optSig}
    The optimal signaling scheme $\pi^*_n$ in $G^{\tau,n}$ has signals space $\SG = [n]$, and satisfies
    \begin{equation}
        y^{\pi^*_n}_i(j)= \begin{cases}
            1 - (n-1)\tau & j = i \\
            \tau & j \neq i.
        \end{cases}
        \quad p^{\pi^*_n}(i) = \frac{1}{n}.
    \end{equation}
\end{fact}
\noindent Fact~\ref{fac:optSig} follows by bounding $BPR(\pi)$ and showing that $\pi^*_n$ attains this value, and we note that the sender can not increase $BPR(\pi)$ with more signals.
When $\tau = \nicefrac{1}{2(n-1)}$, $BPR(\pi^*_n) = \nicefrac{1}{2}$ for all $n$.
We note that even for near-optimal strategies, no signal is sent with probability more than $\calO( \nicefrac{1}{n} + \epsilon)$ when $\tau = \nicefrac{1}{2(n-1)}$.
Indeed, define $\SG^*_\pi$ as the set of signals that induce actions in $\cup_{i\in [n]} \{a_i\}$ in the known-commitment setting, \ie, 
\begin{equation}
    \SG^*_\pi = \{\sg \in \SG | \exists i \in [n] : a_i \in \arg\max_{a\in A} \langle y, \uvec{R}{a} \rangle\}.
\end{equation}
\begin{lemma}
    \label{lem:probBound}
    If $\pi$ is an $\epsilon$-optimal signaling scheme under known-commitment in the flower game $G^{\nicefrac{1}{2(n-1)},n }$, then for all $\sg\in \SG^*_\pi$, $p^\pi(\sg) \leq 2\left( \nicefrac{1}{n} + \epsilon \right)$. 
\end{lemma}
\noindent The proof of Lemma~\ref{lem:probBound} comprises an argument that, for a fixed signal $\sg^*$, near-optimal signaling schemes must place significant probability on signals in $S\setminus \{\sg^*\}$.\iflong\footref{fn:extended_version}\fi

For the flower game,
the posteriors under the optimal scheme are close to distributions under which the receiver takes actions that give the sender no reward, and this closeness makes inferability difficult.
Figure~\ref{fig:rewardViz} shows posteriors under the optimal scheme for $G^{\nicefrac{1}{6},3}$.
The receiver only gets reward when the receiver takes actions in $\cup_{j\in [n]} \{a_j\}$, and the blue inner triangle defines distributions where the receiver takes these actions.
The posteriors under near-optimal schemes will be close to the edges of the blue inner triangle.  
Thus, the receiver's empirical posterior estimate is likely to be in the yellow region, which causes the receiver to take actions
that give no reward to the sender.
Furthermore, as dimension $n$ grows, the probability that the empirical distribution falls outside the blue region grows as, intuitively, the corners of the blue triangle become narrower.

\subsection{Sample complexity lower-bounds}

The receiver must have $\Omega\left( \nicefrac{n}{\epsilon^2(\epsilon + \frac{1}{n})} \right)$ samples before the sender can approximately achieve their known-commitment reward in the inference setting.\iflong\footref{fn:extended_version}\fi
\begin{theorem}
    \label{thm:LBMain}
    For the flower game $G^{\nicefrac{1}{2(n-1)},n}$, let $\pi$ be $\epsilon$-optimal in the known-commitment setting.
    If $IR_k(\pi) \geq \max_{\pi'} BPR(\pi') - \epsilon$, then
    $k \geq {\Omega}\left( \frac{n}{\epsilon^2(\epsilon + \frac{1}{n})} \right)$. 
\end{theorem}
\noindent The $\epsilon + \nicefrac{1}{n}$ term appears via the bound on the probability that each signal is sent.
The $\epsilon^2$ term appears as more-optimal signaling schemes require posteriors closer to the edge of the blue region.
Finally, the $n$ term appears due to changes in the geometry of the blue region with dimension $n$.
For $\epsilon = \nicefrac{1}{n}$, we have a sample complexity lower-bound of $\Omega(n^4)$.

\subsubsection*{Proof roadmap}

We prove Theorem~\ref{thm:LBMain} by showing that, under near-optimal signaling schemes, the process that generates the receiver's samples is close to a process where the receiver gets samples from distributions in the yellow region in Figure~\ref{fig:rewardViz}.
In Lemma~\ref{lem:generalBanditDecompLemma}, we decompose a scheme's inference-setting reward into two terms that relate to the sender's reward when the receiver builds their empirical estimate from two different sample sources.
The first term encodes the distance between the distributions on estimates when the receiver gets samples according to the true posteriors and according to distributions in the yellow regions, respectively.
The second term is the sender's inference setting reward when the receiver builds estimates by sampling from distributions in the yellow region.
We bound the second term as a constant, and we prove Theorem~\ref{thm:LBMain} by bounding the first term as a function of $n$ and $\epsilon$.
Lemma~\ref{lem:averageKLBound} shows that near-optimal schemes induce posteriors that are, on average, close to distributions in the yellow region, and we combine Lemmas~\ref{lem:probBound} and \ref{lem:averageKLBound} to bound the distance term in Lemma~\ref{lem:generalBanditDecompLemma}.

We use techniques from bandit lower bounds by constructing a sequence  $(z_\sg)_{\sg \in \SG}$ of distributions, which we term \emph{failure} distributions, and bounding the reward of a near-optimal scheme $\pi$ by the distance of each posterior $\post{\sg}$ to a failure distribution. 
Lemma~\ref{lem:generalBanditDecompLemma}
codifies this bound, where the first term, in \textcolor{blue}{blue}, is the distance between $z_\sg$ and $\post{\sg}$.
and the second term, in \textcolor{red}{red}, defines the leader's reward when the receiver estimates posteriors under failure distributions.\iflong\footref{fn:extended_version}\fi
\begin{lemma}
    \label{lem:generalBanditDecompLemma}
    Let $\pi$ be a signaling strategy, and $(z_{\sg})_{\sg\in \SG}$ be a tuple of distributions in $\Delta^\Omega$. Additionally, define $\genPostEst{z}{\sg,k}{p} = \frac{w_z}{h}$ for $h \sim 1 + \mathsf{Bin}(k-1, p)$ and $w_z \sim \mathsf{Multi}(h, z_{\sg})$. Then
    \begin{multline}
        \label{eq:bigBanditDecomp}
    IR_k(\pi) \leq 
    \textcolor{blue}{\sqrt{\frac{1}{2}}
        \sum_{\sg \in \SG} \sigProb{\sg} \sum_{a \in A} ( \langle \uvec{S}{a} , \post{\sg} \rangle} \\ \textcolor{blue}{\times \sqrt{(1 + (k-1)\sigProb{\sg})D_{KL}(\genPost{z}{\sg},\post{\sg})} )}
        \\ + \textcolor{red}{\sum_{\sg \in \SG} \sigProb{\sg} \sum_{a \in A} \langle \uvec{S}{a} , \post{\sg} \rangle (\mathbb{P}[\genPostEst{z}{\sg,k}{\sigProb{\sg}} \in E_a])}.
    \end{multline}
\end{lemma}
\noindent Variable $h$ models the number of times the receiver observes a signal, and $w_z$ models the count of each state when they see $h$ samples drawn from the failure distribution $z_s$.

If we choose $(z_s)_{s\in S}$ such that $a(z_s) \notin \bigcup_{j} \{a_j\}$, we can bound the \textcolor{red}{red} term by a constant that is independent of $n$ and strictly less than $\max_{\pi'} BPR(\pi') = \nicefrac{1}{2}$.
We use bounds on the probability of a binomial exceeding its mean \cite{greenberg2014tight} to show that
$\genPostEst{z}{\sg,k}{p} \notin \bigcup_{i\in[n]} E_{a_i}$ with probability at least $\nicefrac{1}{4}$.

With the above bound, it is sufficient to show that 1) no signal has high marginal probability under near-optimal $\pi$, and 2) for near-optimal schemes $\pi$, the expected distance of $\post{\sg}$ to $z_{\sg}$ is small,
for $(z_\sg)_{\sg\in \SG}$ such that $a(z_s) $ is not in $ \smash{\bigcup_{j} \{a_j\}}$.
In particular, when the \textcolor{red}{red} term is strictly bounded away from $\nicefrac{1}{2}$ by a constant, an $f(n,\epsilon)\sqrt{k}$ bound on the \textcolor{blue}{blue} term implies a sample lower bound of $k \geq \Omega\left( \nicefrac{1}{f(n,\epsilon)^2 }\right)$. 
We then approximately bound the \textcolor{blue}{blue} term by
\begin{equation}
    \sqrt{k}\underset{\text{bound using Lemma~\ref{lem:probBound}}}{\left(\sqrt{\max_{\sg \in \SG^*_\pi} \sigProb{\sg}}\right)} \underset{\text{bound using Lemma~\ref{lem:averageKLBound}}}{\left( \sum\nolimits_{\sg \in \SG^*_\pi} \sigProb{\sg} D_{KL}(z_{\sg} ||\post{\sg})\right)}.
\end{equation}

Lemma~\ref{lem:averageKLBound} codifies that posteriors under near-optimal signaling schemes, in expectation, have low KL-divergence with distributions $(z_\sg)_{\sg \in \SG}$ such that $a(z_s)$ is not in $ \bigcup_{j} \{a_j\}$.\iflong\footref{fn:extended_version}\fi
\begin{lemma}
    \label{lem:averageKLBound}
    Let $\pi$ be a signaling strategy such that $\pi$ is $\epsilon$-optimal in the commitment setting.
    Then there exists $(z_{\sg})_{\sg\in \SG^*_\pi}$ such that $\sum_{\sg\in \SG^*_\pi} \sigProb{\sg} \sqrt{D_{KL}(\genPost{z}{\sg}||\post{\sg})} \leq \epsilon\sqrt{\nicefrac{32}{n-1}}$ and, for each $\sg \in \SG^*_\pi$, $\min_i z(i) < \nicefrac{1}{2(n-1)}$.
\end{lemma}
\noindent Lemma~\ref{lem:averageKLBound}'s proof
involves bounding the expected value of $\nicefrac{1}{2} - \max_i \post{\sg}(i)$ by $\epsilon$, where the value $\nicefrac{1}{2} - \max_i \post{\sg}(i)$ serves as a distance between $\post{\sg}$ and a posterior under the optimal scheme.
Intuitively, under an $\epsilon$-optimal strategy, it is impossible for all posteriors to be far from the optimal posteriors.
We then use Lemma~\ref{lem:deltaKLRelation} to construct distributions $(z_{\sg})_{\sg \in \SG}$ with $\calO(\epsilon^2/n)$ KL-divergence to the posteriors $(\post{\sg})_{\sg \in \SG}$.
\begin{lemma}
    \label{lem:deltaKLRelation}
    Let $y \in \mathbb{R}^n$ be such that $y(i) =\frac{1}{2} - \delta$,
    and suppose $a_i \in \arg\max_{a \in A} \langle y, \uvec{R}{a}\rangle$. 
    There exists $z \in \mathbb{R}^n$, $j \in [n]$ such that $D_{KL}(z || y) \leq \nicefrac{32\delta^2}{n-1}$ and $z(j) < \tau$.
\end{lemma}
\noindent The proof of Lemma~\ref{lem:deltaKLRelation} defines $z$ explicitly and uses \cite{dragomir2000some} to upper bound the KL-divergence.
Lemma~\ref{lem:deltaKLRelation} relies on the geometry of the blue inner triangle depicted in Figure~\ref{fig:rewardViz}.

\subsection{Comparing sample complexity against Stackelberg}

We compare the above lower bound to upper bounds on samples required in the Stackelberg setting, to show that Bayesian persuasion problems have higher sample complexity.
For a sequence of Stackelberg games that are geometrically equivalent to $G^{\nicefrac{1}{2(n-1)},n}$, we show that $\calO(n^3 \log (n))$ samples are sufficient to ensure $IR_k(\pi) \geq \max_{x'} SR(x') - \nicefrac{1}{n}$,
where $SR(x)$ is the known-commitment expected value of leader-strategy $x$.
Meanwhile, Theorem~\ref{thm:LBMain} implies that $\Omega(n^4)$ samples are necessary for Bayesian persuasion.

We define a Stackelberg game $G^{\tau,n}_{\mathsf{Stck}}$
which shares geometry with the flower game $G^{\tau,n}$.
The leader's action space $A^L_n$ is $\Omega_n$, i.e., the set of states in $G^{\tau,n}$, while the follower's action space in the Stackelberg game is the receiver's action space in $G^{\tau,n}$.
The utilities in the Stackelberg game satisfy
$
    u_L^{\tau,n}(i,a) = u_S^{\tau,n}(i,a) \text{ and } u_F^{\tau,n}(i,a) = u_L^{\tau,n}(i,a)
$
for $i$ in $A^L_n$ and $a$ in $A_n$.
$G^{\tau,n}_{\mathsf{Stck}}$ and $G^{\tau,n}$ share geometry as each action $a$ is optimal for the same 
distributions on leader actions in $G^{\tau,n}_{\mathsf{Stck}}$ and world states in $G^{\tau,n}$.

The optimal strategy in $G^{\tau,n}_{\mathsf{Stck}}$ is equal to one of the sender's posteriors in $G^{\tau,n}$.
By symmetry, the choice of posterior is arbitrary, and we use $\smash{y_1^{\pi_n^*}}$.
\begin{fact}
    Let $x$ be such that $x(1) = \nicefrac{1}{2}$ and $x(j) = \nicefrac{1}{2(n-1)}$ otherwise. 
    Then $x$ is optimal in $G^{\nicefrac{1}{2(n-1)},n}_{\mathsf{Stck}}$.
\end{fact}
\noindent The follower's optimal action is $a_1$ for strategy $x$.

We give an $\epsilon$-optimal strategy for the leader and give an upper bound on the number of samples required to approximately achieve the leader's known-commitment value.\iflong\footref{fn:extended_version}\fi
\begin{proposition}
    \label{prop:dimTightStackUpperBound}
    Assume $n \geq 4$ and $\epsilon \leq \nicefrac{1}{8}$,
    and let $x_\epsilon$ be the leader strategy in   $G^{\nicefrac{1}{2(n-1)},n}_{\mathsf{Stck}}$ such that $x_\epsilon(1) = \nicefrac{1}{2} - \nicefrac{\epsilon}{2}$ and all other actions are uniform.
    The strategy $x_\epsilon$ satisfies
    $
        IR_k(x_\epsilon) \geq \nicefrac{1}{2} - \epsilon
    $
    if $k \geq \mathcal{O}(\left(\log(n) + \log(\frac{1}{\epsilon})\right)\frac{n}{\epsilon^2})$.
\end{proposition}
\noindent In Proposition~\ref{prop:dimTightStackUpperBound}'s proof, we find a KL-ball around $x_\epsilon$ inside $E_{a_1}$, that we use 
with empirical distribution concentration results \cite{csiszar1984sanov} to bound the probability of
the receiver taking an action in $A\setminus \{a_1\}$.
Compared to upper bounds from \cite{karabag2023should}, this bound has a tighter dependence on $n$ for $G^{\nicefrac{1}{2(n-1)},n}_{\mathsf{Stck}}$.

Proposition~\ref{prop:dimTightStackUpperBound} completes the sample-complexity gap proof.
We emphasize that, as $G^{\tau,n}_{\mathsf{Stck}}$ and  $G^{\tau,n}$ share 
problem geometry, the sample complexity differences appear due to the privileged information unique to Bayesian persuasion.

\section{Designing inferable signaling schemes}

The theoretical results highlight the need to design inferable signaling schemes. 
One could naively pose an optimization problem on the distribution of the receiver's estimates, but
when the number of interactions is high, this optimization problem has size exponential in the number of states.
We propose two approximate methods for designing inferable schemes. The first uses stochastic approximation to maximize $IR_k$, while the second uses an analytical regularization scheme that models the receiver as boundedly-rational.

For both methods, we parameterize the scheme by a joint distribution $X \in \mathbb{R}^{\Omega \times A}$.
We consider schemes where signals are action recommendations, \ie, $S = A$, and signals are persuasive, \ie, signal $a$ induces action $a$.
These schemes are optimal with known-commitment \cite{kamenica2011bayesian}.
We constrain $X$ to lie in the simplex.
Additionally, we ensure consistency with the state distribution, \ie, 
\begin{equation}
    \sum\nolimits_{a'\in A} X(\omega,a') = \mu(\omega) \ \  \forall \omega \in \Omega,
\end{equation}
and persuasiveness of the signals, \ie, 
\begin{equation*}
    \langle  X(:,a), u_R(:,a) \rangle \geq \langle  X(:,a), u_R(:,a') \rangle \ \  \forall a\in A, a'\neq a.
\end{equation*}

\subsection{Projected stochastic gradient descent}
\label{sec:sgd_methods}

While exactly differentiating $IR_k(\pi)$ is hard due to the combinatorial number of receiver estimates, we can simplify the problem by directly estimating the probability with which the receiver takes each action. 
We express $IR_k(\pi)$ as 
\begin{equation}
    IR_k(\pi) = \sum_{\sg \in \SG} \sum_{a \in A} \mathbb{P}[\hat{y}_{\sg,k}^\pi \in E_a ] \langle \uvec{S}{a} , p^\pi(:,s) \rangle.
\end{equation}
Parameterizing $p^\pi(\omega,\sg)$ with $\psgdvar$, the derivative of this expression with respect to $\psgdvar(\hat{\omega},\hat{\sg})$ is
\begin{multline}
    \frac{\partial IR_k(\psgdvar)}{\partial \psgdvar(\hat{\omega},\hat{\sg})} = 
     \sum_{a \in A} \mathbb{P}[\hat{y}_{\hat{\sg},k}^\pi \in E_a ] u_S(\hat{\omega}, a)
     \\ +  \sum_{\sg \in \SG} \sum_{a\in A} \frac{\partial \mathbb{P}[\hat{y}_{\sg,k}^\pi \in E_a ] }{\partial \psgdvar(\hat{\omega},\hat{\sg})} \langle \uvec{S}{a} , \psgdvar(\cdot,\sg) \rangle.
\end{multline}
We estimate $\mathbb{P}[\hat{y}_{\sg,k}^\pi \in E_a ]$ by drawing values of $\hat{y}_{\sg,k}^\pi$.
We then use measure-valued gradients \cite{mohamed2020monte} to estimate the $\nicefrac{\partial \mathbb{P}}{\partial X}$ term by drawing from an augmented distribution.
Let $C([n],l)$ be the set of allocations of $l$ samples to $n$ bins, let $T$ be a subset of $C([n],l)$, and let $Z$ have distribution $\mathsf{Multi}(l,p)$.  %
Define $g_T = \sum_{a \in C([n],l)} \left(\nicefrac{l!}{\prod_i a_i!}\right) \prod_{i} p_i^{a_i}1_T(a)$.
For $p \in \Delta^{[n]}$, $g_T$ is the probability that $Z$ falls in $T$.
For $p \in \Delta^{[n]}$, we can evaluate gradients of $g_T$ by sampling according to
\begin{equation}
    \frac{\partial g_T(p)}{\partial p_i} = l\mathbb{P}_{\hat{Z} \sim \mathsf{Multi}(k-1,p)}[\hat{Z} + \mathbf{e}_i \in T].
\end{equation}
We can use similar logic to evaluate $\frac{\partial \mathbb{P}[\hat{y}_{\sg',k}^\pi \in E_a ] }{\partial \psgdvar(\omega,\sg)}$.
\iflong\footref{fn:extended_version}\fi

We apply this estimator in projected stochastic gradient descent \cite{garrigos2023handbook} to maximize $IR_k(\pi)$ for given $k$,
though the method also applies to cumulative reward $\sum_{j=1}^k IR_j(\pi)$.

\subsection{Regularized information design}
\label{sec:reg_methods}

Stochastic gradient descent requires extensive sampling, so we next provide a sample-free method for Problem~\ref{prob:optProblem}.

The receiver's inference induces randomness in their actions, and we model this randomness with bounded rationality  \cite{mckelvey1995quantal}.
For a rationality constant $\lambda \in [0,\infty)$, a receiver with posterior $\post{\sg}$ on the state will take action $a$ with probability
\begin{equation}
    p_\lambda(a; \post{\sg}) = \frac{ \exp\left(\lambda \sum_{\omega \in \Omega} u_R(\omega,a) \post{\sg}(\omega)\right) }{ \sum_{a \in A} \exp\left(\lambda \sum_{\omega \in \Omega} u_R(\omega,a) \post{\sg}(\omega)\right)}.
\end{equation}
When $\lambda = 0$, the receiver's actions are uniformly random.
The sender's optimization problem is then
\begin{equation}
    \label{eq:brrProblem}
    \max_\pi \mathbb{E}_{\omega \sim \mu, \sg \sim \pi(\omega)}[\mathbb{E}_{a \sim p_\lambda(a; \post{\sg})}[u_S(\omega,a)]].
\end{equation}
\noindent The bounded rationality model smooths the sender's reward landscape, as posteriors near the receiver's decision boundary cause more random receiver actions.
Thus, schemes solving \eqref{eq:brrProblem} move posteriors away from decision boundaries, and this movement increases inferability, as codified in Proposition~\ref{prop:upperBoundMain}.
We solve \eqref{eq:brrProblem} via projected gradient descent with a decaying step size \cite{garrigos2023handbook}.
While we synthesize schemes with a bounded-rationality receiver model, 
we evaluate schemes for a receiver who acts rationally with respect to their posterior estimates.

\section{Numerical examples}

\subsection{Evaluating SGD and regularization in the flower game}
\begin{figure}
    \centering
    \includegraphics[width=0.98\linewidth]{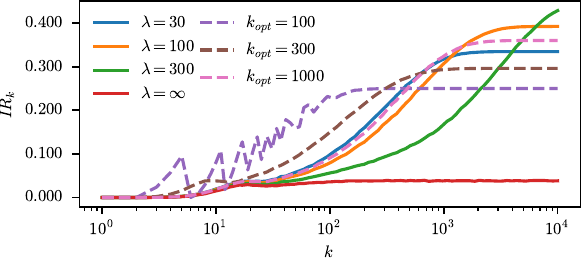}
    \caption{\textbf{Schemes found with SGD outperform regularization for small $k$, while for high $k$, we can tune performance through the rationality constant.} 
    We plot estimates of $IR_k$ for policies found using each method for the flower game.
    Dashed lines correspond to SGD, where we optimize $IR_{k_{opt}}$.
    Solid lines correspond to regularization with different $\lambda$ values.
    We evaluate SGD schemes averaged over the last $10$ SGD iterates, and we tune step sizes as different $k_{opt}$ and $\lambda$ values lead to different smoothness properties.
    SGD schemes perform best up to $k \approx 300$.
    Beyond $k = 300$, regularization performs well, and we can tune $\lambda$ to trade long and short-term performance.
    The known-commitment solution, \ie, $\lambda= \infty$, performs poorly as posteriors lie on receiver decision boundaries, and thus, for any $k$, 
    the receiver takes poor actions for the sender with high probability.
    } 
    \label{fig:sgd_br_comp}
\end{figure}

\begin{figure}
    \centering
    \includegraphics[width=0.98\linewidth]{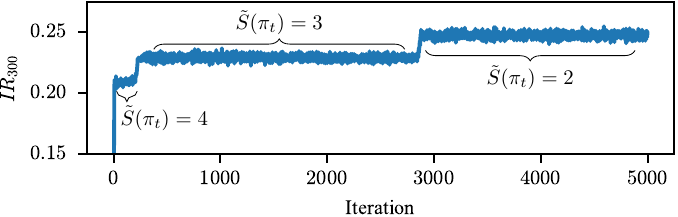}
    \caption{\textbf{Optimal schemes in the inferability setting have small signal spaces.} 
    We plot $IR_{300}(\pi_t)$ when optimizing $IR_{300}$ with SGD, where $\pi_t$ is the  $t^{\text{th}}$ iterate.
    Define $\tilde{S}(\pi)$ as the size of the smallest set $\hat{S}$ of signals so that $\sum_{s\in \hat{S}} \sigProb{\sg} \langle \uvec{S}{a^\pi(\sg)}, \post{\sg}\rangle $ is at least $0.99 BPR(\pi)$, i.e., the number of signals the receiver must learn for the sender to recover $BPR(\pi)$.
    We mark the values of $\tilde{S}(\pi_t)$.
    SGD improves the scheme by reducing $\tilde{S}(\pi)$.}
    \label{fig:sgd_natural_sig_red}
\end{figure}

We compare stochastic gradient descent (SGD) and regularization approaches for the flower game $G^{\nicefrac{1}{6},4}$, and
Figure~\ref{fig:sgd_br_comp} shows that the schemes found with SGD outperform regularization when the receiver has few interaction-samples.
One explanation for this observation is that SGD can find signaling schemes where the receiver needs to learn fewer signals, which improves inferability in the short run at the expense of long-run performance, and  Figure~\ref{fig:sgd_natural_sig_red} validates that SGD finds schemes with smaller signal spaces.

Figure~\ref{fig:sgd_br_comp} also shows that, in the high-interaction regime, optimizing for bounded-rationality receivers provides a tunable method for inferable persuasion, where
decreasing $\lambda$ trades long-term performance for short-term gains.
Indeed, for small $k$, the receiver is less predictable due to limited interaction samples, and low rationality constants model this behavior better.
We note that, in the flower game, 
very low rationality constants, \ie, below $\lambda = 30$, significantly change the value landscape in \eqref{eq:brrProblem}, and thus break the above trend.

Both approaches outperform the known-commitment schemes, as known-commitment-optimal schemes place posteriors at receiver decision boundaries.

\subsection{Evaluating regularization on random games}

We generate $100$ random persuasion problems with $|\Omega| = |A| = 4$ and uniform state distributions, where $u_R(\omega,a)$ and $u_S(\omega,a)$ are drawn uniformly from $[0,1]$. 
We apply regularization for different values of $\lambda$ to these games.
We do not apply SGD in this setting due to computational concerns.

\begin{figure}
    \centering
    \includegraphics[width=0.98\linewidth]{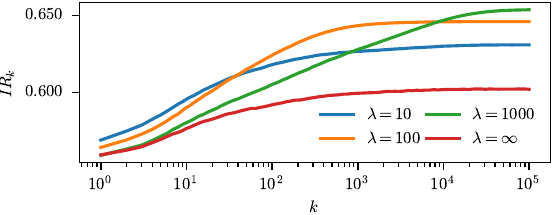}
    \caption{\textbf{Optimizing for boundedly-rational receivers provides a flexible regularization framework for the inference setting.}
    We plot estimates of $IR_k$ for schemes derived with various $\lambda$ values on random games. 
    We average over the approximately $60$ games on which projected gradient descent converges for all $\lambda$, and we tune step-sizes for different $\lambda$ values.
    Decreasing $\lambda$ trades long-term performance for gains when $k$ is small.
    Known-commitment-optimal schemes, \ie, $\lambda = \infty$, show poor performance for all $k$ values as posteriors may sit at decision boundaries.
    }
    \label{fig:br_reg}
\end{figure}

Figure~\ref{fig:br_reg} confirms that bounded-rationality regularization provides a tunable proxy for optimizing $IR_k$, and that these schemes outperform the known-commitment baseline.

\subsection{Evaluating SGD in the safety-alert example}

We revisit the example from the introduction where a planner, i.e., a sender, designs alerts for people visiting a city, i.e., the receiver. 
On each day, the planner observes an incident that affects certain city blocks as their privileged information and signals people entering the city. 
The planner aims to keep people far from the true incident location. 
However, people want to visit locations close to the city center while avoiding the incident. 
The planner and people have partially aligned goals, as the people do not want to visit the incident site.
However, goals may conflict, as only the planner cares about the distance of people from the incident.

We model this setting as a Bayesian persuasion problem.
The city is a weighted graph $(V,E,W)$ where $v^*$ in $V$ represents the city center.
Each state $\omega$ is an incident, corresponding to a set $F_\omega \subset V$ of affected nodes.
The distribution of incidents is uniform.
The receiver's action is a node $v \in V$ to visit, and they get negative reward proportional to the distance between $v$ and $v^*$ if $v \notin F_\omega$, and a constant penalty if $v \in F_\omega$.
The sender's reward is proportional to the minimum distance of $v$ to a node in $F_\omega$.
We measure distances according to the weights $W$.

\begin{figure}
    \centering
    \includegraphics[width=0.8\linewidth]{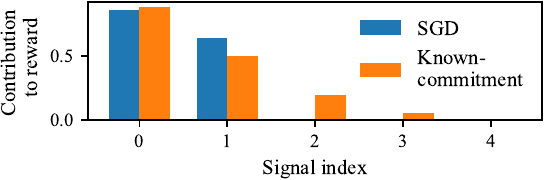}
    \caption{\textbf{The optimal scheme in the safety example relies on a smaller signal size than the known-commitment-optimal solution.}
    For policies $\pi$ derived by optimizing known-commitment reward, and by solving the SGD problem for $k = 100$, we plot the value
    $\sum_{\omega \in \Omega }p^\pi(\omega, \sg) u_S(\omega, a^\pi(\sg))$, i.e., the contribution of signal $\sg$ to the planner's total expected value.
    Under the SGD-derived scheme, the planner's reward concentrates in fewer signals.
    }
    \label{fig:reward_decomp}
\end{figure}

\begin{figure}
    \centering
    \includegraphics[width=0.73\linewidth]{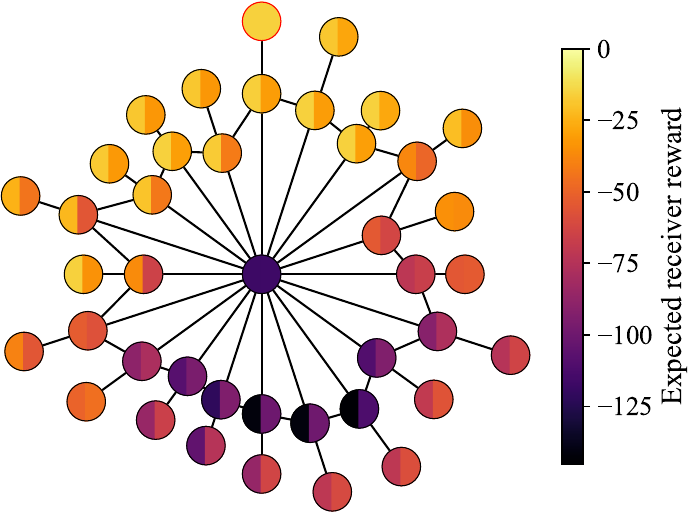}
    \caption{\textbf{The optimal action in the safety example is more distinct under the scheme derived with SGD.}
    Let $\pi_{kc}$ be the optimal known-commitment scheme, and let $\pi_{sgd}$ be the scheme we find with SGD.
    For the first signal $\sg^*$ shown in Figure~\ref{fig:reward_decomp}, and for each node $v$, we color the left half by expected reward under posterior $y^{\pi_{kc}}_{\sg^*}$ and the right half by the expected reward under posterior $y^{\pi_{sgd}}_{\sg^*}$.
    Under the SGD-derived scheme, the people's optimal action, bordered in red, is more distinct.
    }
    \label{fig:post_map_example}
\end{figure}

Figures~\ref{fig:reward_decomp} and \ref{fig:post_map_example} show that, when compared to the known-commitment-optimal scheme $\pi_{kc}$  the SGD-derived scheme $\pi_{sgd}$ has a smaller signal space and makes the receiver's action more distinct. 
We solve an alert example 
with $|V| = 40$, $|F_\omega| = 10$ and $|\Omega| = 20$, for $k = 100$.
The value $BPR(\pi_{sgd})$ for the SGD-derived scheme $\pi_{sgd}$ comprises contributions for two signals, compared to four for $\pi_{kc}$.
Furthermore, for posteriors that $\pi_{sgd}$ induces, the second-best action for the receiver is better separated when compared to the posteriors that scheme $\pi_{kc}$ induces.


\section{Conclusion}
We explored Bayesian persuasion in settings where receivers infer a signaling scheme from samples.
We provided theoretical characterizations of the receiver's reward in the inference settings as a function of the signaling scheme.
We then proposed two methods for synthesizing inferable signaling schemes, one using stochastic approximation and the other using regularization.
We showed both methods to improve upon the baseline of applying known-commitment-optimal schemes in the inference setting.

\bibliographystyle{plain} %
\bibliography{references}

\begin{thebibliography}{10}

\bibitem{an2012security}
Bo~An, David Kempe, Christopher Kiekintveld, Eric Shieh, Satinder Singh, Milind Tambe, and Yevgeniy Vorobeychik.
\newblock Security games with limited surveillance.
\newblock In {\em AAAI Conference on Artificial Intelligence}, volume~26, pages 1241--1248, 2012.

\bibitem{assos2024maximizing}
Angelos Assos, Yuval Dagan, and Constantinos Daskalakis.
\newblock Maximizing utility in multi-agent environments by anticipating the behavior of other learners.
\newblock In {\em Advances in Neural Information Processing Systems}, volume~37, pages 38769--38798, 2024.

\bibitem{best2024persuasion}
James Best and Daniel Quigley.
\newblock Persuasion for the long run.
\newblock {\em Journal of Political Economy}, 132(5):1740--1791, 2024.

\bibitem{blum2014lazy}
Avrim Blum, Nika Haghtalab, and Ariel Procaccia.
\newblock Lazy defenders are almost optimal against diligent attackers.
\newblock In {\em AAAI Conference on Artificial Intelligence}, volume~28, pages 573--579, 2014.

\bibitem{camara2022mechanism}
Modibo~K Camara.
\newblock Mechanism design with a common dataset.
\newblock {\em EC}, page 558, 2022.

\bibitem{camara2020mechanisms}
Modibo~K Camara, Jason~D Hartline, and Aleck Johnsen.
\newblock Mechanisms for a no-regret agent: Beyond the common prior.
\newblock In {\em {IEEE} annual symposium on foundations of computer science}, pages 259--270. IEEE, 2020.

\bibitem{chao1972negative}
Min-Te Chao and WE~Strawderman.
\newblock Negative moments of positive random variables.
\newblock {\em Journal of the American Statistical Association}, 67(338):429--431, 1972.

\bibitem{chen2025explainable}
Yiling Chen, Tao Lin, Wei Tang, and Jamie Tucker-Foltz.
\newblock Explainable information design.
\newblock {\em arXiv:2508.14196}, 2025.

\bibitem{collina2024efficient}
Natalie Collina, Aaron Roth, and Han Shao.
\newblock Efficient prior-free mechanisms for no-regret agents.
\newblock In {\em ACM Conference on Economics and Computation}, pages 511--541, 2024.

\bibitem{conitzer2006computing}
Vincent Conitzer and Tuomas Sandholm.
\newblock Computing the optimal strategy to commit to.
\newblock In {\em ACM conference on Electronic commerce}, pages 82--90, 2006.

\bibitem{cover1999elements}
Thomas~M Cover.
\newblock {\em Elements of information theory}.
\newblock John Wiley \& Sons, 2006.

\bibitem{csiszar1984sanov}
Imre Csisz{\'a}r.
\newblock Sanov property, generalized {I}-projection and a conditional limit theorem.
\newblock {\em The Annals of Probability}, 12(3):768--793, 1984.

\bibitem{das2017reducing}
Sanmay Das, Emir Kamenica, and Renee Mirka.
\newblock Reducing congestion through information design.
\newblock In {\em Allerton conference on communication, control, and computing}, pages 1279--1284, 2017.

\bibitem{dong2022optimal}
Hongcheng Dong and Yifen Mu.
\newblock The optimal strategy against fictitious play in infinitely repeated games.
\newblock In {\em Chinese Control Conference (CCC)}, pages 6852--6857. IEEE, 2022.

\bibitem{dragomir2000some}
Sever~S Dragomir, Marcel~L Scholz, and Jadranka Sunde.
\newblock Some upper bounds for relative entropy and applications.
\newblock {\em Computers \& Mathematics with Applications}, 39(9-10):91--100, 2000.

\bibitem{dughmi2016persuasion}
Shaddin Dughmi, David Kempe, and Ruixin Qiang.
\newblock Persuasion with limited communication.
\newblock In {\em ACM Conference on Economics and Computation}, pages 663--680, 2016.

\bibitem{dughmi2016algorithmic}
Shaddin Dughmi and Haifeng Xu.
\newblock Algorithmic {B}ayesian persuasion.
\newblock In {\em ACM symposium on Theory of Computing}, pages 412--425, 2016.

\bibitem{feng2024rationality}
Yiding Feng, Chien-Ju Ho, and Wei Tang.
\newblock Rationality-robust information design: {B}ayesian persuasion under quantal response.
\newblock In {\em Annual ACM-SIAM Symposium on Discrete Algorithms}, pages 501--546. SIAM, 2024.

\bibitem{fudenberg1998theory}
Drew Fudenberg and David~K Levine.
\newblock {\em The theory of learning in games}.
\newblock MIT press, 1998.

\bibitem{garrigos2023handbook}
Guillaume Garrigos and Robert~M Gower.
\newblock Handbook of convergence theorems for (stochastic) gradient methods.
\newblock {\em arXiv:2301.11235}, 2023.

\bibitem{greenberg2014tight}
Spencer Greenberg and Mehryar Mohri.
\newblock Tight lower bound on the probability of a binomial exceeding its expectation.
\newblock {\em Statistics \& Probability Letters}, 86:91--98, 2014.

\bibitem{jain2024calibrated}
Atulya Jain and Vianney Perchet.
\newblock Calibrated forecasting and persuasion.
\newblock {\em arXiv:2406.15680}, 2024.

\bibitem{kamenica2011bayesian}
Emir Kamenica and Matthew Gentzkow.
\newblock Bayesian persuasion.
\newblock {\em American Economic Review}, 101(6):2590--2615, 2011.

\bibitem{karabag2023should}
Mustafa~O Karabag, Sophia Smith, Negar Mehr, David Fridovich-Keil, and Ufuk Topcu.
\newblock When should a leader act suboptimally? the role of inferability in repeated {S}tackelberg games.
\newblock {\em arXiv:2310.00468}, 2023.

\bibitem{le2019persuasion}
Ma{\"e}l Le~Treust and Tristan Tomala.
\newblock Persuasion with limited communication capacity.
\newblock {\em Journal of Economic Theory}, 184:104940, 2019.

\bibitem{lin2024generalized}
Tao Lin and Yiling Chen.
\newblock Generalized principal-agent problem with a learning agent.
\newblock {\em arXiv:2402.09721}, 2024.

\bibitem{lin2024information}
Tao Lin and Ce~Li.
\newblock Information design with unknown prior.
\newblock {\em arXiv:2410.05533}, 2024.

\bibitem{mckelvey1995quantal}
Richard~D McKelvey and Thomas~R Palfrey.
\newblock Quantal response equilibria for normal form games.
\newblock {\em Games and economic behavior}, 10(1):6--38, 1995.

\bibitem{min2021bayesian}
Daehong Min.
\newblock Bayesian persuasion under partial commitment.
\newblock {\em Economic Theory}, 72:743--764, 2021.

\bibitem{mohamed2020monte}
Shakir Mohamed, Mihaela Rosca, Michael Figurnov, and Andriy Mnih.
\newblock Monte {C}arlo gradient estimation in machine learning.
\newblock {\em Journal of Machine Learning Research}, 21:1--62, 2020.

\bibitem{muthukumar2020learning}
Vidya~K Muthukumar.
\newblock {\em Learning from an unknown environment}.
\newblock PhD thesis, University of California, Berkeley, 2020.

\bibitem{peng2019bayesian}
Cheng Peng and Masayoshi Tomizuka.
\newblock Bayesian persuasive driving.
\newblock In {\em American Control Conference}, pages 723--729. IEEE, 2019.

\bibitem{tang2021bayesian}
Wei Tang and Chien-Ju Ho.
\newblock On the {B}ayesian rational assumption in information design.
\newblock In {\em AAAI Conference on Human Computation and Crowdsourcing}, volume~9, pages 120--130, 2021.

\bibitem{vundurthy2023intelligent}
Bhaskar Vundurthy, Aris Kanellopoulos, Vijay Gupta, and Kyriakos~G Vamvoudakis.
\newblock Intelligent players in a fictitious play framework.
\newblock {\em IEEE Transactions on Automatic Control}, 69(1):479--486, 2024.

\bibitem{yang2024computational}
Kunhe Yang and Hanrui Zhang.
\newblock Computational aspects of {B}ayesian persuasion under approximate best response.
\newblock {\em Advances in Neural Information Processing Systems}, 37:134430--134458, 2024.

\bibitem{zu2021learning}
You Zu, Krishnamurthy Iyer, and Haifeng Xu.
\newblock Learning to persuade on the fly: Robustness against ignorance.
\newblock In {\em ACM Conference on Economics and Computation}, pages 927--928, 2021.

\end{thebibliography}

\addtolength{\textheight}{-12cm}   %

\ifappfn
\addtolength{\textheight}{12cm}
\clearpage
\section*{APPENDIX}

\section{Expanded proofs}

\begin{proof_alt}[Proof of Proposition~\ref{prop:upperBoundMain}]
We compute the difference in the round-$k$ reward by decomposing according to the signal sent. 
Let $\omega,\sg,\hat{a}$ be distributed as
\begin{equation*}
     \omega \sim \mu, 
    \quad \sg \sim \pi(\omega), 
    \quad a \sim \actDist{\sg}.
\end{equation*}
In this case, we bound the gap as
\begin{multline*}
    \mathbb{E}_{\omega, \sg}[u_S(\omega,a^\pi(\sg)) ] - \mathbb{E}_{\omega, \sg, \hat{a}}[u_S(\omega,\hat{a})]
    \\ = \sum_{\sg'\in \SG} \mathbb{E}_{\omega, \sg,\hat{a}}\left[(u_S(\omega,a^\pi(\sg)) - u_S(\omega,\hat{a}(\sg))) 1_{\sg = \sg'} 1_{\hat{a}(\sg) \neq a(\sg) }\right]
    \\ \leq \sum_{\sg' \in \SG} \mathbb{E}_{\omega, \sg}[ 1_{\sg = \sg'} 1_{\hat{a} \neq a(\sg) }] = \sum_{\sg} \sigProb{\sg} \mathbb{P}_{\hat{a} \sim \actDist{s}}(\hat{a} \neq a(\sg)).
\end{multline*}
In the inequality, we use the boundedness of utilities, and in the final equality, we use the fact that the receiver action $\hat{a}$ is conditionally independent of the signal.

After invoking the Markov bound from \cite{karabag2023should}, we have
\begin{equation}
     \sum_{\sg} \sigProb{\sg} \mathbb{P}(\hat{a} \neq a(\sg)) \leq \sum_{\sg} \sigProb{\sg} \frac{\mathbb{E}[||\postEst{\sg,k} - \post{\sg}||_2]}{d(\post{\sg})},
\end{equation}
and it remains to bound $\mathbb{E}[||\postEst{\sg,k} - \post{\sg}||_2]$.

We compute $\mathbb{E}[||\postEst{\sg,k} - \post{\sg}||_2]$ by conditioning on the number of times the receiver observes signal $\sg$ before round $k$, \ie
\begin{equation}
    \sum_{h = 1}^k \mathbb{E}\left[||\postEst{\sg,k} - \post{\sg} ||_2 \ | \  \#(\sg) = h\right] \mathbb{P}(\#(\sg) = h).
\end{equation}

Conditional on $\# (\sg) = h$, we can bound  $\mathbb{E}[||\postEst{\sg,k} - \post{\sg} ||_2|]$ by 
$\nicefrac{\nu( \post{\sg})}{\sqrt{h}}$, using Lemma 1 from \cite{karabag2023should}, and so we bound the total expectation as
\begin{multline}
    \label{eq:upperBoundPreInvBinEval}
    \mathbb{E}[||\postEst{\sg,k} - \post{\sg} ||_2] \leq \sum_{h = 1}^k {\frac{\nu( \post{\sg})}{\sqrt{h}}} P(\#(\sg) = h)
    \\ %
    \leq {\nu( \post{\sg})} \sqrt{\sum_{h=1}^k P(\# (s) = h) \frac{1}{h}},
\end{multline}
where we apply Jensen's inequality for the last step.

To bound the  term in the root, we note that $\# (s) \sim 1 + \mathsf{Bin}(k-1, p^\pi(\sg))$, and we apply the expression for negative moments of binomial distribution from \cite{chao1972negative} to bound \eqref{eq:upperBoundPreInvBinEval} by
\begin{equation}
\nu( \post{\sg}) \sqrt{\frac{1 - (1 - p^\pi(\sg))^k}{k p^\pi(\sg)}}
    \\ \leq  \nu( \post{\sg}) \sqrt{\frac{1}{k \sigProb{\sg}}}.
\end{equation}
\end{proof_alt}

\begin{proof_alt}[Proof of Corollary~\ref{cor:MIBound}]
    For $x$ a distribution on states, by convexity of $-\log$ we have
    \begin{equation}
        \nu^2(x) \leq
        H(x).
    \end{equation}
    Appealing to Proposition~\ref{prop:upperBoundMain} and applying H\"older's inequality with the vectors $\mathbf{1}$ and $(\sqrt{\sigProb{\sg}H(\post{\sg})})_{\sg\in \SG}$ then gives
    \begin{equation}
        \mathsf{GAP}_\pi \leq \sqrt{|\SG|} \sqrt{\sum_{\sg \in \SG}\sigProb{\sg} H(\post{\sg})}.
    \end{equation}
\end{proof_alt}

\begin{proof_alt}[Proof of Lemma~\ref{lem:actOptCharac}]
    The $\implies$ direction follows by checking that $a_i$ gives higher utility than every action.
    
    ($\impliedby$) $\langle \uvec{R}{a_i}, y\rangle \geq \langle \uvec{R}{a_j}, y\rangle$ for all $j \neq i$ follows from condition 1), and the fact that
    $\langle \uvec{R}{a_i}, y\rangle \geq \langle \uvec{R}{a_{(ij)}}, y\rangle$ for all $j \neq i$ follows from condition $2$. We next need to show that, for $i\neq j \neq k$ and $i \neq k$, we have $\langle \uvec{R}{a_i}, y\rangle \geq \langle \uvec{R}{a_{(jk)}}, y\rangle$. 
    This statement is equivalent to 
    $
        y_i \geq y_j + \tau - y_k,
    $
    and this fact follows from conditions 1) and 2).
    Finally, we need to show $\langle \uvec{R}{a_i}, y\rangle \geq \langle \uvec{R}{a_{(ji)}}, y\rangle$ for $i \neq j$, however, conditions 1) and 2) imply that $y_i \geq \tau$ also, and so we conclude that this statement holds.
\end{proof_alt}

\begin{proof_alt}[Proof of Lemma~\ref{lem:probBound}]
    Let $p^\pi(\cdot,\cdot)$ denote the joint distribution on world states and signals. 
    For a fixed signal $\sg^*$ we aim to bound $p^\pi(\sg^*) = \sum_{\omega \in \Omega} p^\pi (\omega, \sg^*)$.
    As $p$ is a distribution, we have
    \begin{equation}
        \sum_{\omega \in \Omega} p^\pi(\omega,\sg^*) \leq 1 - \sum_{\omega \in \Omega, \sg \in \SG \setminus \{\sg^*\}} p^\pi(\omega,\sg).
    \end{equation}

    For signal $\sg \in \SG^*_\pi$, let $\omega_s$ be some element in $\arg\max_{\omega \in \Omega} p^\pi(\omega,\sg)$. 
    We remark that the reward of $\pi$ in the commitment setting is $\sum_{\sg \in \SG^*_\pi}  p^\pi(\omega_s,\sg)$.
    Thus, by the $\epsilon$-optimality assumption on $\pi$,
    \begin{equation}
        \sum_{\sg \in \SG^*_\pi \setminus \{\sg^*\} } p^\pi(\omega_s,\sg) \geq \frac{1}{2} - \epsilon - p^\pi(\omega_{\sg^*},\sg^*)
    \end{equation}

    For a signal $\sg$ in $\SG^*_\pi$, let $\omega^\#_{\sg}$ be an element in $\arg\min_{\omega\in \Omega} p^\pi(\omega,\sg)$.
    Using Lemma~\ref{lem:actOptCharac}, we have 
    have that 
    \begin{multline}
        p^\pi(\omega^\#_{\sg},\sg) \geq \tau \sum_{\omega \in \Omega} p^\pi(\omega,\sg) \\ \geq \tau( p^\pi(\omega_s,\sg) + (n-1)p^\pi(\omega^\#_{\sg},\sg) ),
    \end{multline}
    and rearranging, we have, for $\tau = \frac{1}{2(n-1)}$, that
    \begin{equation}
        p^\pi(\omega^\#,\sg) \geq \frac{\tau}{1 - (n-1)\tau} p^\pi(\omega_s,\sg) = \frac{1}{n-1} p^\pi(\omega_s,\sg).
    \end{equation}
    We now deduce a number of facts about the optimal signaling scheme.
    We first apply the above inequality to deduce the following.
    \begin{multline}
        \sum_{\omega \in \Omega} p^\pi(\omega,\sg^*) \leq 1 - \sum_{\omega \in \Omega, \sg \in \SG^*_\pi \setminus \{\sg^*\}} p^\pi(\omega,\sg) \\ \leq 1 - 2 \sum_{\sg \in \SG^*_\pi \setminus \{\sg^*\} } p^\pi(\omega_s,\sg)  \\ \leq 1 - 2(\nicefrac{1}{2} - \epsilon) + 2p^\pi(\omega_{\sg^*},\sg^*) = 2\epsilon + 2p^\pi(\omega_{\sg^*},\sg^*).
    \end{multline}
    Finally, we note that, for all $\omega$, we must have $\sum_{\sg \in \SG} p^\pi(\omega,\sg) = \frac{1}{n}$, due to the fact that the $\omega$-marginal of $p^\pi$ must be $\mu$.
    Thus we deduce $p^\pi(\omega_{\sg^*},\sg^*) \leq \frac{1}{n}$,
    and so finally, we conclude 
    \begin{equation}
        \sum_{\omega \in \Omega} p^\pi(\omega,\sg^*) \leq 2\left( \epsilon + \frac{1}{n} \right). 
    \end{equation}
\end{proof_alt}

\begin{proof_alt}[Proof of Lemma~\ref{lem:generalBanditDecompLemma}]
    We can expand $IR_k(\pi)$ as follows
    \begin{multline}
        IR_k(\pi) = \sum_{\sg \in \SG} \sigProb{\sg} \sum_{a \in A} \langle \uvec{S}{a} , \post{\sg} \rangle \mathbb{P}[\postEst{\sg,k} \in E_a]
        \\= 
        \sum_{\sg \in \SG} \sigProb{\sg} \ldots \\ \times \sum_{a \in A} \langle \uvec{S}{a} , \post{\sg} \rangle (\mathbb{P}[\postEst{\sg,k} \in E_a] - \mathbb{P}[\genPostEst{z}{\sg,k}{\sigProb{\sg}} \in E_a])
        \\ + \sum_{\sg \in \SG} \sigProb{\sg} \sum_{a \in A} \langle \uvec{S}{a} , \post{\sg} \rangle (\mathbb{P}[\genPostEst{z}{\sg,k}{\sigProb{\sg}} \in E_a]),
    \end{multline}
    We can upper bound the difference in probabilities by the KL-divergence between the distributions that define $\postEst{\sg,k}$ and $\genPostEst{z}{\sg,k}{p(\sg)}$ using Pinsker's inequality.
    Before evaluating this divergence we first note that $\postEst{\sg,k}$ is distributed according to 
    \begin{multline}
        h \sim 1 + \mathsf{Bin}(k-1, \sigProb{\sg}), \\ w_y \sim \mathsf{Multi}(h, \post{\sg}), \quad \postEst{\sg,k} = \frac{w_y}{h_y}.
    \end{multline}
    We define $C(\Omega,l)$ as the set of allocations of $l$ samples to $\Omega$.
    We then bound the KL-divergence as follows.
    \begin{multline}
        D_{KL}(\calD_{\genPostEst{z}{\sg,k}{\sigProb{\sg}}},\mathcal{D}_{\postEst{\sg,k}}) 
        \leq D_{KL}(\calD_{w_z,h},\mathcal{D}_{w_y,h})
        \\ = \sum_{h = 1}^k p(\#(\sg) = h) \sum_{a \in C(\Omega, h)} \\ \times\mathbb{P}_{\mathsf{Multi}(h,\genPost{z}{\sg})}(a) \log\left(\frac{\mathbb{P}_{\mathsf{Multi}(h,\genPost{z}{\sg})}(a)}{\mathbb{P}_{\mathsf{Multi}(h,\post{\sg})}(a)} \right)
        \\ = \sum_{h = 1}^k p(\#(\sg) = h) h D_{KL}(\genPost{z}{\sg},\post{\sg})
        \\ = (1 + (k-1)\sigProb{\sg}) D_{KL}(\genPost{z}{\sg},\post{\sg}).
    \end{multline}
    The first inequality applies data-processing.
    The second equality expands the probabilities of all cases for the counts.
    The third equality follows by applying the binomial probability distribution of $\#(s)$.
\end{proof_alt}

\begin{proof_alt}[Proof of Lemma~\ref{lem:deltaKLRelation}]
    If $y(i) = \frac{1}{2} - \delta$, we can deduce $\sum_{j \neq i} y(j) \leq \frac{1}{2} + \delta$ and thus, there exists $j$ such that $y(j) \leq \frac{1}{2(n-1)} + \frac{\delta}{(n-1)}$.
    Let $\tau' = \max(\frac{1}{4(n-1)},\frac{1}{2(n-1)} - \frac{\delta}{n-1})$ and note that $\tau' \in (0,\tau)$.
    We define $z$ with
    \begin{equation}
        z(i) = \begin{cases}
              \tau' & i = j \\
             y(i) + \frac{y(j) - \tau'}{n-1} & i \neq j.
        \end{cases}
    \end{equation}
    
    We can use \cite{dragomir2000some} to bound the KL-divergence between these two distributions. Indeed, for each $i $ we have
    $
        z(i) -  y(i) 
        \leq \frac{2\delta}{(n-1)^2},
    $
    and thus we have, for each $i\neq j$
    \begin{equation}
        \frac{z(i)}{y(i)}
        \leq \frac{2\delta}{y(i) (n-1)^2} +1 
        \leq \frac{4\delta}{ (n-1) } +1,
    \end{equation}
    where in the second equality we apply the fact that $a_i \in \arg \max_{a\in A} \langle y , \uvec{R}{a} \rangle$, and so Lemma~\ref{lem:actOptCharac} implies $y(i) \geq \tau = \frac{1}{2(n-1)}$.
    Meanwhile, we have
    \begin{multline}
        \frac{z(i)}{y(i)} \geq \frac{\tau'}{ \tau' + y(j) - \tau'} \geq \frac{\tau'}{\tau' + \frac{2\delta}{n-1}} = \\ \frac{1}{ 1 + \frac{2\delta}{\tau' (n-1)}} \geq 1 - \frac{2\delta}{\tau' (n-1)} \geq 1 - 8\delta.
    \end{multline}
    The third inequality uses the fact that, for $x > 0$, $\frac{1}{1+x} > 1-x$.
    Finally, we can deduce the lemma using Theorem 6 of \cite{dragomir2000some}.
\end{proof_alt}

\begin{proof_alt}[Proof of Lemma~\ref{lem:averageKLBound}]
    Given a signal $\sg \in \SG^*_\pi$, we define
    \begin{equation}
        \delta_s^\pi = \frac{1}{2} - \max_{i} \post{\sg}(i). 
    \end{equation}
    Informally, $\delta_s^\pi$ quantifies how close a posterior is to a posterior under the optimal known-commitment scheme $\pi^*$, and, through Lemma~\ref{lem:deltaKLRelation}, bounds on $\delta_s^\pi$ facilitate bounds on $KL$-divergence of posteriors.
    
    If $p^\pi(\omega,\sg)$ is the joint distribution on states and signals, then the quantity $\sigProb{\sg} \delta_s^\pi$ is equal to
    \begin{equation}
        \label{eq:jointDistToGap}
        \frac{1}{2} \sum_{\omega\in \Omega} p^\pi(\omega,s) - p(i^*_{\sg},\omega),
    \end{equation}
    where $i^*_{\sg} \in \Omega$ is an index which attains the maximum in $\post{\sg}$. 
    When we sum \eqref{eq:jointDistToGap} over $\SG^*_\pi$, we have
    \begin{equation}
        \label{eq:totalPDelValue}
        \sum_{\sg\in \SG^*_\pi} \sigProb{\sg}  \delta_s^\pi = \frac{1}{2} \sum_{\sg\in \SG^*_\pi, \omega\in \Omega} p^\pi(\omega,\sg) - \sum_{\sg\in \SG^*_\pi}p^\pi(i^*_{\sg},\sg).
    \end{equation}
    We can upper bound the first term by $\nicefrac{1}{2}$ as $p^\pi(\cdot,\cdot)$ is a distribution and we also note that
    \begin{equation}
        p^\pi(i^*_{\sg},\sg) = \sigProb{\sg}  y_s^\pi(i^*_{\sg}) = \sigProb{\sg}  \langle 
        \uvec{S}{a^\pi(\sg)} , y_s^\pi \rangle.
    \end{equation}
    The second equality uses Lemma~\ref{lem:actOptCharac}.
    In the full-observability setting, the sender only gets reward from signals in $\SG^*_\pi$, and thus  
    \begin{equation}
     BPR(\pi) = \sum_{\sg\in \SG^*} \sigProb{\sg}  \langle \uvec{S}{a^\pi(\sg)} , y_s^\pi \rangle = \sum_{\sg\in \SG^*_\pi}p^\pi(i^*_{\sg},\sg).
    \end{equation}
    Using the assumption that $BPR(\pi) \geq \nicefrac{1}{2} - \epsilon$ with \eqref{eq:totalPDelValue} we have
    \begin{equation}
        \sum_{\sg\in \SG^*_\pi} \sigProb{\sg}  \delta_s^\pi \leq  \frac{1}{2} - \left( \frac{1}{2} - \epsilon \right) = \epsilon.
    \end{equation}
    We can then apply Lemma~\ref{lem:deltaKLRelation} to conclude that we can construct $(z_s)_{\sg \in \SG}$ such that the stated result holds.
\end{proof_alt}

\begin{proof_alt}[Proof of Theorem~\ref{thm:LBMain}]
    Let $\pi$ be such that $BPR(\pi) \geq \frac{1}{2} - \epsilon$.

    For $\sg \in \SG^*_\pi$, we set define $z_s$ according to Lemma~\ref{lem:averageKLBound}, so that $\sum_{\sg\in \SG^*} \sigProb{\sg} \sqrt{D_{KL}(z_s||\post{\sg})} \leq \epsilon \sqrt{\frac{32}{n-1}}$, and for each posterior $z_s$ there exists a state with probability strictly smaller than $\tau$.
    For $\sg \in \SG \setminus \SG^*_\pi$, we set $z_s = \post{\sg}$, and note that these $z_s$ also must have an element smaller than $\tau$.

    Applying Lemma~\ref{lem:generalBanditDecompLemma}, we have
    \begin{multline}
        \label{eq:mainThmMainDecomp}
        IR_k(\pi) \\
        \leq \sqrt{\frac{1}{2}} \sum_{\sg \in \SG} p^\pi(\sg) \sqrt{(1 + (k-1)p^\pi(\sg)) D_{KL}(y^\pi_s,z_s)}
        \\\cdot \sum_{a \in A} \langle \uvec{S}{a},\post{\sg} \rangle
        \\ + \sum_{\sg \in \SG} p^\pi(\sg) \sum_{a\in A} \langle \uvec{S}{a}, \post{\sg} \rangle \mathbb{P}[\genPostEst{z}{\sg,k}{p(\sg)} \in E_a].
    \end{multline}

    We first note that, by definition of $u_S$ and $A$,
    \begin{equation}
        \sum_{a \in A} \langle \uvec{S}{a},\post{\sg} \rangle = \sum_{i \in [n]} \langle \mathbf{e}_i, \post{\sg} \rangle = 1.
    \end{equation}

    We bound the first term in \eqref{eq:mainThmMainDecomp} using the properties of $z_s$ we construct in Lemma~\ref{lem:deltaKLRelation}. 
    We have
    \begin{multline}
        \sum_{\sg \in \SG^*_\pi} p^\pi(\sg) \sqrt{(1 + (k-1)p^\pi(\sg)) D_{KL}(y^\pi_s,z_s)}
        \\ \leq \sum_{\sg \in \SG^*_\pi} p^\pi(\sg) (1 + \sqrt{(k-1)p^\pi(\sg)}) \sqrt{D_{KL}(y^\pi_s,z_s)}
        \\ \leq  \sum_{\sg \in \SG^*_\pi} p^\pi(\sg) \sqrt{D_{KL}(y^\pi_s,z_s)}  \\ + \sqrt{k-1} \sum_{\sg \in \SG^*_\pi} (p^\pi(\sg))^{1.5} \sqrt{D_{KL}(y^\pi_s,z_s)} 
        \\ \leq   \sum_{\sg \in \SG^*_\pi} p^\pi(\sg) \sqrt{D_{KL}(y^\pi_s,z_s)} \\ + \sqrt{k-1} \max_{\sg \in \SG^*_\pi} \sqrt{p^\pi(\sg)} \sum_{\sg \in \SG^*_\pi} p^\pi(\sg) \sqrt{D_{KL}(y^\pi_s,z_s)} 
        \\ \leq \frac{\sqrt{32}}{\sqrt{n-1}} \left( \epsilon + 2 \sqrt{k-1} \sqrt{\epsilon + \frac{1}{n}} \epsilon \right).
    \end{multline}
    We drop signals in $S\setminus S^*_\pi$ as they satisfy $D_{KL}(\post{\sg},z_{\sg}) = 0$
    The second equality uses $\sqrt{1 + x} \leq 1 + \sqrt{x}$ for $x \in [0,\infty) $. 
    The final equality applied Lemma~\ref{lem:averageKLBound} and \ref{lem:probBound}.

    It remains to bound
    \begin{equation}
        \label{eq:infResidual}
        \sum_{\sg \in \SG} p^\pi(\sg) \sum_{a\in A} \langle \uvec{S}{a}, \post{\sg} \rangle \mathbb{P}[\genPostEst{z}{\sg,k}{\sigProb{\sg}} \in E_a]
    \end{equation}

    We have
    \begin{equation}
        BPR(\pi) \leq \sum_{\sg\in \SG^*_\pi} p^\pi(\sg) \langle \uvec{S}{a^\pi(\sg)}, \post{\sg}\rangle,
    \end{equation}
    as the leader only gets value from signals in $\SG^*$ in the known-commitment setting, and for signals in $\SG^*$, we note that $$\langle \uvec{S}{a^\pi(\sg)}, \post{\sg}\rangle \leq \frac{1}{2}$$, which we can deduce from Lemma~\ref{lem:actOptCharac}. Thus, we have
    \begin{equation}
        \frac{1}{2} - \epsilon \leq BPR(\pi) \leq \frac{1}{2} \sum_{\sg\in \SG^*_\pi} p^\pi(\sg),
    \end{equation}
    which implies
    \begin{equation}
        \sum_{\sg\notin \SG^*_\pi} p^\pi(\sg) \leq 2\epsilon.
    \end{equation}
    We then bound the portion of \eqref{eq:infResidual} corresponding to $\SG^*_\pi$ with
    \begin{multline}
        \sum_{\sg\in \SG^*_\pi} \sigProb{\sg} \sum_{i \in [n] } \post{\sg}(i) \mathbb{P}[\genPostEst{z}{\sg,k}{\sigProb{\sg}} \in E_{a_i}],
        \\\leq \frac{1}{2} \sum_{\sg\in \SG^*_\pi} \sigProb{\sg} \sum_{i \in [n]} \mathbb{P}[\genPostEst{z}{\sg,k}{\sigProb{\sg}} \in E_{a_i}]
        \\\leq \frac{1}{2} \max_{\sg\in \SG^*_\pi} \left( \sum_{i \in [n]} \mathbb{P}[\genPostEst{z}{\sg,k}{\sigProb{\sg}} \in E_{a_i}] \right)
    \end{multline}

    We now claim that $\sum_i \mathbb{P}[\genPostEst{z}{\sg,k}{\sigProb{\sg}} \in E_{a_i}] \leq \frac{3}{4}$ for all $\sg \in \SG^*_\pi$.
    If $\# \sg =1$, $\genPostEst{z}{\sg,k}{\sigProb{\sg}}$ lies at a corner of the simplex, and thus
    $\mathbb{P}[\genPostEst{z}{\sg,k}{\sigProb{\sg}} \in E_{a_i}| \#(\sg) = 1] = 0$.
    If $\# \sg = 2$, we can apply Theorem 1 of \cite{greenberg2014tight}.
    Indeed, by assumption, for a given $z_s$, there exists $i$ such that $z_s(i) < \tau$, and thus for $v \sim \mathsf{Multi}( \# \sg, z_\sg)$, we have
    \begin{multline}
        \mathbb{P}\left[\sum_{j \in [n] \setminus \{i\}} v(j) \geq \# \sg (1 - z_s(i))\right] > \frac{1}{4},
        \\ \implies \mathbb{P}[v(i) \leq \# \sg z_s(i)] > \frac{1}{4}
        \\ \implies \mathbb{P}[\genPostEst{z}{\sg,k}{p^\pi(\sg)}(i) < \tau| \# \geq 2] > \frac{1}{4}
        \\ \implies \sum_{i \in [n] }\mathbb{P}[\genPostEst{z}{\sg,k}{p^\pi(\sg)} \in E_{a_i}] \leq \frac{3}{4},
    \end{multline}
    where in the final step we apply Lemma~\ref{lem:actOptCharac}.
    Thus, we conclude that
    \begin{equation}
        \sum_{\sg \in \SG} p^\pi(\sg) \sum_{a\in A} \langle \uvec{S}{a}, \post{\sg} \rangle \mathbb{P}[\genPostEst{z}{\sg,k}{p(\sg)} \in E_a] \leq 2\epsilon + \frac{3}{8}.
    \end{equation}
    We can now bound the reward in the inferability setting as
    \begin{equation}
        IR_k(\pi) \leq \frac{D}{\sqrt{n-1}} \left( \epsilon + 2 \sqrt{k-1} \sqrt{\epsilon + \frac{1}{n}} \epsilon \right) + 2 \epsilon + \frac{3}{8},
    \end{equation}
    where $D = 4$.
    We aim to compute the resulting lower bound on $k$ so that $IR_k(\pi) \geq \frac{1}{2} - \epsilon$. We thus have
    \begin{equation}
        \frac{2D}{\sqrt{n-1}}\sqrt{k-1} \sqrt{\epsilon + \frac{1}{n}}\epsilon \geq \frac{1}{8} - \frac{D}{\sqrt{n-1}}\epsilon - 3 \epsilon,
    \end{equation}
    and for $\epsilon \leq \frac{1}{64(D+1)}, n\geq 2$, this implies
    \begin{equation}
        \frac{2D}{\sqrt{n-1}}\sqrt{k-1} \sqrt{\epsilon + \frac{1}{n}} \epsilon \geq \frac{1}{16},
    \end{equation}
    and rearranging, we get
    \begin{equation}
        k \geq \frac{1}{1024 D^2} (n-1) \frac{1}{\epsilon^2(\epsilon + \frac{1}{n})}
    \end{equation}
\end{proof_alt}

\begin{proof_alt}[Proof of Proposition~\ref{prop:dimTightStackUpperBound}]
    Define $x'_\delta \in \Delta^{[n]}$ such that $x'_\delta(1) = \nicefrac{1}{2} - \delta$, and all other elements are equal.
    We construct a KL-ball around $x'_\delta$ that $E_{a_1}$ contains, \ie,
    we construct $t$ such that $y \notin E_{a_1}$ implies $D_{KL}(y||x'_\delta) > t$.

    We lower bound $D_{KL}(y||x'_\delta)$ for $y$ such that $\rho = \min_{i \neq 1} y(i) \leq \tau$.
    Indeed, we have, by data processing, that
    \begin{multline}
        D_{KL}\left(y||x'_\delta\right) \geq D_{KL,2}\left(\rho||\frac{1}{2(n-1)} + \frac{\delta}{n-1}\right)
        \\ \geq D_{KL,2}\left(\frac{1}{2(n-1)}||\frac{1}{2(n-1)} + \frac{\delta}{n-1}\right),
    \end{multline}
    where $D_{KL,2}(a,b)$ is the KL-divergence between Bernoulli distributions with parameters $a$ and $b$ respectively.

    We then note that
    \begin{equation}
        D_{KL,2}(p||p + \gamma) \geq \frac{1}{4}\frac{\gamma^2}{p},
    \end{equation}
    for $\frac{\gamma}{p} \in [0,1]$,
    which follows by finding appropriate quadratic lower bounds on the terms comprising $D_{KL,2}$.
    Thus, we can deduce that
    \begin{equation}
        D_{KL}(y||x'_\delta) \geq \frac{1}{2} \frac{\delta^2}{n-1}
    \end{equation}
    Thus we can deduce that $\min_i y(i) \leq \tau$ implies $D_{KL}(y||x'_\delta) \geq \frac{1}{2} \frac{\delta^2}{n-1}$.
    Using Pinsker's inequality, we can also deduce, that, in the ball $\mathcal{B}_{KL}(x'_\delta, \delta^2/(2(n-1)) ) $, $y(1)$ is the largest element. Indeed, for $n \geq 4, \delta \leq \frac{1}{8}$ and $y$ in the ball we have
    \begin{equation}
        y(1) \geq \frac{1}{2} - \delta - \delta / (2\sqrt{n-1}) \geq \frac{3}{8} - \frac{1}{16} = \frac{5}{16},
    \end{equation}
    and, for $j\neq 1$,
    \begin{equation}
        y(j) \leq \frac{\frac{1}{2} + \delta}{n-1} + \delta / (2\sqrt{n-1})
        \\ \leq \frac{5/8}{3} + \frac{1}{16} < \frac{5}{16}.
    \end{equation}
    By the characterization of $E_{a_1}$ in Lemma~\ref{lem:actOptCharac}, we can use a union bound to deduce
    \begin{multline}
        \mathbb{P}[\hat{(x'_\delta)}_k \notin E_{a_1}] \leq \sum_{i \in [n]} \mathbb{P}[\hat{(x'_\delta)}_k(i) \leq \tau] \\
        + \sum_{i \in [n] \setminus \{1\}} \mathbb{P}[\hat{(x'_\delta)}_k(i) \geq \hat{(x'_\delta)}_k(1)].
    \end{multline}
    All the sets $C$ that define the above probability bound are convex, and in fact, half-spaces, and thus we can bound the probabilities as
    \begin{equation}
        \mathbb{P}[{(\hat{x'}_\delta)}_k \in C] \leq e^{-kD_{KL}(C||x'_\delta)},
    \end{equation}
    where $D_{KL}(C||x'_\delta)$ is the minimum KL-divergence to $x'_\delta$ \cite{csiszar1984sanov}.
    For all the composite sets, the minimum KL divergence is at least $\frac{1}{2} \frac{\delta^2}{n-1}$, and thus we can conclude that
    \begin{equation}
        \mathbb{P}[{(\hat{x'}_\delta)}_k \notin E_{a_1}] \leq 2n \exp\left(-k \frac{1}{2} \frac{\delta^2}{n-1}\right),
    \end{equation}
    and applying this bound to $x_\epsilon = x'_{\epsilon/2}$ we have
    \begin{equation}
        \mathbb{P}[{(\hat{x}_\epsilon)}_k \notin E_{a_1}] \leq 2n \exp\left(-k \frac{1}{8} \frac{\epsilon^2}{n-1}\right).
    \end{equation}
    To ensure $IR_{k}(x_\epsilon) \geq \frac{1}{2} - \epsilon$, it is sufficient to ensure $\mathbb{P}[\hat{(x_\epsilon)}_k \notin E_{a_1}]$ is smaller than $\frac{\epsilon}{2}$, \ie
    \begin{equation}
        2n \exp\left(-k \frac{1}{8} \frac{\epsilon^2}{n-1}\right) \leq \epsilon/2.
    \end{equation}
    Rearranging we obtain the following sufficient condition on $k$
    \begin{equation}
        k\geq 8\left( \log(2n) + \log\left(\frac{2}{\epsilon}\right)\right) \frac{n}{\epsilon^2}
    \end{equation}
\end{proof_alt}

\section{Gradient estimation details}

In this section, we derive an expression for 
$\frac{\partial \mathbb{P}[\postEst{\sg,k+1} \in E_a]}{\partial X(\omega,s)}$.
We give results for $k+1$ to make this derivation cleaner.
When $p^\pi = X$, we recall that we model the receiver's counts at time $k+1$ by
\begin{multline}
    \#(\omega,\sg)_{k+1} = Z(\omega,\sg) + Y_{\sg}(\omega) : \\
     Z \sim \mathsf{Multi}(k,X) \quad Y_{\sg} \sim \mathsf{Multi}\left( 1,\left( \frac{X(\omega,s)}{\sum_{\omega' \in \Omega}X(\omega',s)}\right)_{\omega \in \Omega}\right).
\end{multline}

Define $T_{s,a}$ as the set of values for $Z$ and $Y_s$ such that $\postEst{\sg}$ induces action $a$. 
\begin{multline*}
    \left\{ c,\omega_0 | \left( \frac{c(\omega,s) + 1_{\omega_0}(\omega)}{\sum_{\omega} c(\omega,s) + 1}\right)_{\omega \in \Omega} \in E_a  \right\},
\end{multline*}
where $(c,\omega_0)$ lies in  $C(\Omega \times S,k) \times \Omega$. 
Note that $T_{s,a}$ does not depend on $Y_{s'}$ for $s' \neq s$.

We aim to estimate
\begin{equation}
    \mathbb{P}[(Z,Y_s) \in T_{s,a}] = \mathbb{E}[1_{T_{s,a}}(Z,Y_s)].
\end{equation}

We can expand this probability as
\begin{multline}
    g(X) = \\
    \sum_{c, \omega_0 } k! \prod_{\omega,s} \frac{X(\omega,s)^{c(\omega,s)}}{c(\omega,s) !} \cdot \frac{X(\omega,s)}{\sum_{\omega'} X(s,\omega')} 1_{T_{s,a}} (c, \omega_0).
\end{multline}
We aim to evaluate $\partial g / \partial X(\hat{\omega},\hat{s})$, and we give this value as two separate components.

We first evaluate the term corresponding to differentiating through the multinomial.
\begin{multline}
    \sum_{c, \omega_0 } k! \prod_{\omega,s} \frac{\partial}{\partial X(\hat{\omega},\hat{s})} \left( \frac{X(\omega,s)^{c(\omega,s)}}{c(\omega,s) !} \right) \\ \times \frac{X(\omega,s)}{\sum_{\omega'} X(s,\omega')} 1_{T_{s,a}} (c, \omega_0)
    \\ = \sum_{c: c(\hat{\omega},\hat{s}) > 0, \omega_0 } k! \frac{X(\hat{\omega},\hat{s})^{c(\hat{\omega},\hat{s}) - 1}}{c(\hat{\omega},\hat{s}) -1 !} \prod_{\omega,s \neq \hat{\omega},\hat{s}} \left( \frac{X(\omega,s)^{c(\omega,s)}}{c(\omega,s) !} \right) \\\times \frac{X(\omega,s)}{\sum_{\omega'} X(s,\omega')} 1_{T_{s,a}} (c, \omega_0)
    \\ = k \sum_{c: c(\hat{\omega},\hat{s}) > 0, \omega_0 } \mathbb{P}_{\tilde{Z}\sim \mathsf{Multi}(k-1, X)} (\tilde{Z} = c - 1_{\hat{\omega},\hat{s}})\\\times \frac{X(\omega,s)}{\sum_{\omega'} X(s,\omega')} 1_{T_{s,a}} (c, \omega_0)
    \\ = k\mathbb{P}_{\tilde{Z} \sim \mathsf{Multi}(k-1, X), Y_s }[(\tilde{Z} + 1_{\hat{\omega},\hat{s}},Y_s) \in T_{s,a}].
\end{multline}
We can easily evaluate this final term via sampling.

We then evaluate the term corresponding to differentiating through the categorical variable with distribution $s$.
\begin{multline}
    \frac{\partial}{\partial X(\hat{\omega},\hat{s})}\left( \frac{X(\omega_0,s)}{\sum_{\omega'} X(\omega',s)} \right)\\ = \begin{cases}
        \frac{1}{\sum_{\omega'} X(\omega',s)} - \frac{X(\omega_0,s)}{(\sum_{\omega'} X(\omega',s))^2} & s = \hat{s}, \hat{\omega} = \omega_0 
        \\ - \frac{X(\omega_0,s)}{(\sum_{\omega'} X(\omega',s))^2} & s = \hat{s}, \hat{\omega} \neq \omega_0
        \\ 0 & \text{otherwise}
    \end{cases}
\end{multline}
Thus we can rewrite the derivative of the second term as 
\begin{multline}
    \sum_{c, \omega_0 } k! \prod_{\omega,s} \frac{X(\omega,s)^{c(\omega,s)}}{c(\omega,s) !} \\ \times \frac{\partial}{\partial X(\hat{\omega},\hat{s})}\left( \frac{X(\omega_0,s)}{\sum_{\omega'} X(\omega',s)} \right) 1_{T_{s,a}} (c, \omega_0)
    \\ = 1_{s = \hat{s}} \sum_{\omega_0} \left( 1_{\omega_0 = \hat{\omega}} \frac{1}{\sum_{\omega'} X(\omega',s)} - \frac{X(\omega_0,s)}{(\sum_{\omega'} X(\omega',s))^2}\right) \\ \times \sum_{c} k! \prod_{\omega,s} \frac{X(\omega,s)^{c(\omega,s)}}{c(\omega,s) !} 1_{T_{s,a}}(c,\omega_0)
    \\ = 1_{s = \hat{s}} \sum_{\omega_0} \left( 1_{\omega_0 = \hat{\omega}} \frac{1}{\sum_{\omega'} X(\omega',s)} - \frac{X(\omega_0,s)}{(\sum_{\omega'} X(\omega',s))^2}\right) \\ \times \mathbb{P}_Z[(Z,\omega_0) \in T_{s,a}].
\end{multline}
We can again easily evaluate this term by sampling $Z$.

Thus, we can evaluate $\partial g / \partial X$ as 
\begin{multline}
    \frac{\partial g(X)}{\partial X(\hat{\omega},\hat{s})} 
    \\ = k\mathbb{P}_{\tilde{Z} \sim \mathsf{Multi}(k-1, X), Y_s }[(\tilde{Z} + 1_{\hat{\omega},\hat{s}},Y_s) \in T_{s,a}]
    \\ + 1_{s = \hat{s}} \sum_{\omega_0 \in \Omega} \left( 1_{\omega_0 = \hat{\omega}} \frac{1}{\sum_{\omega'} X(\omega',s)} - \frac{X(\omega_0,s)}{(\sum_{\omega'} X(\omega',s))^2}\right) \\ \times \mathbb{P}_Z[(Z,\omega_0) \in T_{s,a}].
\end{multline}
We remark that this equality holds when $X$ lies in the simplex $\Delta^{\Omega \times S}$, but we note that we always apply a projection onto the simplex before evaluating this gradient when applying stochastic gradient descent.

\fi

\end{document}